\documentclass[useAMS]{mn2e}
\usepackage{graphicx}
\usepackage{epsfig}
\usepackage{amssymb}
\usepackage{lscape}
\usepackage{ulem}
\usepackage{txfonts}

\def\logms{$\log$ (M$_{\star}$/M$_{\odot}$)}
\def\aq{ALMaQUEST}

\def\sigh2{$\Sigma_{\rm H_2}$}
\def\sighi{$\Sigma_{\rm HI}$}
\def\sigsfr{$\Sigma_{\rm SFR}$}

\def\sigstar{$\Sigma_{\star}$}

\def\kms{km~s$^{-1}$}

\def\c2s{C\,{\sc ii}$^{\star}$}

\def\fgas{$f_{\rm H_2}$}

\title[ALMaQUEST V - Resolved scaling relations] {The ALMaQUEST Survey: V.  The non-universality of kpc-scale star formation relations and the factors that drive them}

\author[Ellison et al.] {Sara L. Ellison$^1$,   Lihwai Lin$^2$, Mallory D. Thorp$^1$, Hsi-An Pan$^3$, Jillian M. Scudder$^4$, \newauthor Sebastian F. S\'{a}nchez$^5$, Asa F. L. Bluck$^6$, Roberto Maiolino$^6$\\ 
$^1$ Department of Physics \& Astronomy, University of Victoria, Finnerty Road, Victoria, British Columbia, 
  V8P 1A1, Canada\\
  $^2$ Institute of Astronomy \& Astrophysics, Academia Sinica, Taipei 10617, Taiwan\\
  $^3$ Max-Planck-Institut f\"ur Astronomie, K\"onigstuhl 17, D-69117 Heidelberg, Germany\\
  $^4$ Department of Physics and Astronomy, Oberlin College, Oberlin, Ohio, OH 44074, USA\\
  $^5$ Instituto de Astronom\'{i}a, Universidad Nacional Autonoma de Mexico, A. P. 70-264, C.P. 04510, Mexico, 
  D.F., Mexico\\
  $^6$ Kavli Institute for Cosmology \& Cavendish Astrophysics, University of Cambridge, Madingley Road, Cambridge, CB3 0HA, UK\\
}

\begin{document}

\maketitle

\begin{abstract}
Using a sample of $\sim$15,000 kpc-scale star-forming spaxels in 28 galaxies drawn from the ALMA-MaNGA QUEnching and STar formation (\aq) survey, we investigate the galaxy-to-galaxy variation of the `resolved' Schmidt-Kennicutt relation (rSK; \sigh2\ - \sigsfr), the `resolved' star forming main sequence (rSFMS; \sigstar\ - \sigsfr) and the `resolved' molecular gas main sequence (rMGMS; \sigstar\ - \sigh2).   The rSK relation, rSFMS and rMGMS all show significant galaxy-to-galaxy variation in both shape and normalization, indicating that none of these relations is universal between galaxies.  The rSFMS shows the largest galaxy-to-galaxy variation and the rMGMS the least.  By defining an `offset' from the average relations, we compute a $\Delta$rSK, $\Delta$rSFMS, $\Delta$rMGMS for each galaxy, to investigate correlations with global properties.  We find the following correlations with at least 2$\sigma$ significance: the rSK is lower (i.e. lower star formation efficiency) in galaxies with higher M$_{\star}$, larger Sersic index and lower specific SFR (sSFR);  the rSFMS is lower (i.e. lower sSFR) in galaxies with higher M$_{\star}$ and larger Sersic index;  the rMGMS is lower (i.e. lower gas fraction) in galaxies with lower sSFR.   In the ensemble of all 15,000 data points, the rSK relation and rMGMS show equally tight scatters and strong correlation coefficients, compared with a larger scatter and weaker correlation in the rSFMS.   Moreover, whilst there is no correlation between $\Delta$rSK and $\Delta$rMGMS in the sample, the offset of a galaxy's rSFMS does correlate with both of the other two offsets.  Our results therefore indicate that the rSK and rMGMS are independent relations, whereas the rSFMS is a result of their combination.
\end{abstract}


\section{Introduction}

In science, observing correlations between macroscopic variables has the potential to convey vital insight into underlying physical processes.  There are numerous examples of such scaling relations in the field of extra-galactic astronomy, such as the Tully-Fisher relation (Tully \& Fisher 1977), the Faber-Jackson relation (Faber \& Jackson 1976), the M-$\sigma$ relation (Ferrarese \& Merritt 2000) and the luminosity- or mass-metallicity relation (Lequeux et al. 1979).  However, disentangling the fundamental nature of such relations from the inter-relation between variables can be complex and leads to the frequent caveat that `correlation does not imply causation'.  Extracting meaningful physical insight from empirical correlations therefore requires careful multi-variate dissection in order to identify the underlying drivers of these correlations (e.g. Bothwell et al. 2016; Dey et al. 2019; Bluck et al 2020a,b).

In galaxy evolution, the present day gas reservoir, the current star formation rate (SFR) and the total stellar mass (M$_{\star}$) of a galaxy encapsulate the on-going and cumulative history of the galaxy's growth.  For star forming galaxies\footnote{In this paper, we focus on the nature of star-forming galaxies and kpc-scale spaxels.  Galaxies and regions that are not actively star forming, or ionized by different processes, are known to deviate from the standard scaling relations (e.g. as reviewed by Sanchez 2020) and are not considered here.  However, see Ellison et al (2021) for an analysis of the resolved molecular gas main sequence in `retired' spaxels in \aq.}, these three global parameters show strong inter-correlations.  A galaxy's total SFR correlates almost linearly with total stellar mass in what has become known as the `star-forming main sequence' (SFMS; Brinchmann et al. 2004; Daddi et al. 2007; Noeske et al. 2007a; Salim et al. 2007).  In turn, the star formation rate correlates with total (Schmidt 1959; Kennicutt 1989), molecular (Wong \& Blitz 2002; de los Reyes \& Kennicutt 2019) and dense molecular (Gao \& Solomon 2004; Wu et al. 2005) gas.   The latter of these has been proposed to be the most `fundamental' since the correlation between star formation and dense molecular gas extends from scales of individual clouds up to galaxies with relatively little scatter (e.g. Lada et al. 2010, 2012).  Finally, the molecular and atomic gas masses correlate with global galaxy stellar mass, as does the fraction of gas in the molecular phase (Saintonge et al. 2011a, 2017; Huang et al. 2012; Bothwell et al. 2014; Cicone et al. 2017; Catinella et al. 2010, 2018; Calette et al. 2018; Sorai et al. 2019; Casasola et al. 2020; Hunt et al. 2020).  Unsurprisingly then, all three of these star formation related variables (SFR, M$_{\star}$ and gas fractions) have been found to be inter-connected, with gas content playing an important role in regulating the global SFMS (Tacconi et al. 2013, 2018; Saintonge et al. 2016, 2017; Piotrowska et al. 2020; Colombo et al. 2020).

Despite the existence of these global galaxy scaling relations between SFR, stellar mass and gas, the underlying process that they trace (i.e. star formation) occurs on much smaller scales.  A deeper insight into the fundamental physics of the star formation process, and principles that regulate it, will therefore be best revealed by investigating scaling relations on the sub-galactic scale.  For the remainder of this paper, following the convention that has emerged in the literature, we will refer to such sub-galactic (typically, kpc-scale) scaling relations as `resolved'\footnote{In practice, the resolution at which galactic scaling relations have been studied ranges from tens of parsecs to a few kpc.  The actual process of star formation is not actually resolved in even the highest angular resolution extra-galactic studies.}.

The kpc-scale correlation between the molecular gas surface density (\sigh2) and the SFR surface density (\sigsfr), also known as the resolved Schmidt-Kennicutt (rSK) relation, is one of the most extensively studied resolved galaxy scaling relations.  Local samples of up to a few tens of galaxies that combine CO, HI and broad band (e.g. optical and/or infra-red)  data have been used to study the interplay of molecular and atomic gas with star formation at kpc or sub-kpc scales (e.g. Wong \& Blitz 2002; Bigiel et al. 2008; Schruba et al. 2011; Leroy et al. 2013).  Several important results have emerged from these detailed local galaxy studies.  First, galaxies show little correlation between \sigsfr\ and \sighi\ (but see Bacchini et al. 2019 for a \textit{volumetric} correlation between HI and SFR).  The early correlations found between total gas surface densities and \sigsfr\ (e.g. Kennicutt 1989, 1998a,b) are entirely driven by a tight sequence between \sigh2\ and \sigsfr\ (e.g. Wong \& Blitz 2002; Bigiel et al. 2008).  However, the correlation between \sigh2\ and \sigsfr\ (i.e. the rSK relation) shows significant galaxy-to-galaxy variation (as well as variation within a given galaxy), that dominates the scatter in the ensemble data (e.g. Bigiel et al. 2008; Schruba et al. 2011;  Leroy et al. 2013; Shetty, Kelly \& Bigiel 2013; Casasola et al. 2015; Zabel et al. 2020).  The variable rSK relations can be re-cast as galaxy-to-galaxy variations in star formation efficiency (SFE=\sigsfr/\sigh2) which are found to have higher values in lower mass, low metallicity galaxies (e.g. Schruba et al. 2011; Leroy et al. 2013; Utomo et al. 2018) and in late morphological types (e.g. Colombo et al. 2018; Sanchez 2020). 

The advent of large optical integral field unit (IFU) surveys, which enable kpc scale measurements of \sigsfr\ and \sigstar, quickly led to the demonstration that the global SFMS is driven by a local scale relationship of stellar mass and SFR surface densities (e.g. Sanchez et al. 2013; Cano-Diaz et al. 2016, 2019; Gonzalez-Delgado et al. 2016; Hsieh et al. 2017; Ellison et al. 2018; Medling et al. 2018), although the presence of a kpc-scale relation between SFR and stellar mass had been `discovered' some two decades earlier (e.g. Ryder \& Dopita 1994).  Thanks to the abundance of data now available in IFU surveys for these two variables, the resolved SFMS (rSFMS) is perhaps the most extensively studied of the three relations considered in this paper. Like the rSK relation, there is significant galaxy-to-galaxy variation seen in the kpc-scale rSFMS (e.g. Abdurro'uf \& Akiyama 2017; Vulcani et al. 2019, 2020; Enia et al. 2020; Hemmati et al. 2020; Casasola et al. in prep).  Morphology seems to play an important role in the regulation of the rSFMS for a given galaxy (Gonzalez-Delgado et al. 2016; Maragkoudakis et al. 2017; Medling et al. 2018; Pan et al. 2018; Cano-Diaz et al. 2019), as is also well established to be true for the global SFMS (e.g. Wuyts et al. 2011; Morselli et al. 2017; Cano-Diaz et al. 2019; Cook et al. 2020; Sanchez 2020; Leslie et al. 2020).      However, when the sample is limited to disk galaxies, the rSFMS seems to be less variable, with no residual mass dependence (Erroz-Ferrer et al. 2019; Enia et al. 2020).   Other factors which may affect an individual galaxy's rSFMS include total stellar mass (Belfiore et al. 2018; Wang et al. 2019), the presence of an active galactic nucleus (AGN; Cano-Diaz et al. 2016, 2019; Sanchez et al. 2018), a recent interaction (e.g. Pan et al. 2019; Thorp et al. 2019) and environment (Schaefer et al. 2017; Medling et al. 2018; Vulcani et al. 2020; Zabel et al. 2020).

Less well-studied is the resolved molecular gas main sequence (rMGMS) correlation between \sigstar\ and \sigh2.  Radial profiles of \sigh2\ have been presented in numerous works, demonstrating that high \sigh2\ occurs at small radii, which coincide with high \sigstar\ (e.g. Leroy et al. 2008; Schruba et al. 2011; Bigiel \& Blitz 2012).   Explicit correlations between \sigstar\ and \sigh2\ were then shown by Shi et al. (2011) for global quantities and Wong et al. (2013) for galactic sub-regions.    Recently, Lin et al. (2019) showed that the rMGMS shows a scatter that is equally as tight as the rSK relation (see also Morselli et al. 2020) and proposed that the existence of a rMGMS reflected the importance of the stellar potential in governing interstellar medium (ISM) conditions.  Although reminiscent of the idea of an `extended' SK relation, in which \sigsfr\ $\propto$ \sigh2$^{\alpha}$\sigstar$^{\beta}$ (e.g. Shi et al. 2011), Lin et al. (2019) did not find a significant reduction in the scatter around the rSK relation when \sigstar\ was included, concluding that the rSK and rMGMS are independently regulated.  Finally, in contrast to studies of the resolved rSFMS and rSK relation, there has been no work to date on whether the rMGMS (whose ensemble scatter is small, Lin et al. 2019; Morselli et al. 2020) might exhibit galaxy-to-galaxy variations.

We will collectively refer to the rSFMS, the rMGMS and the rSK relation as the ‘star formation scaling relations’ in this paper.  In spite of the abundance of study on these scaling relations, past studies have generally tackled only one (or at most, two) of these scaling relations at a time.  Our goal in this paper is to present a unified analysis of the three resolved star formation scaling relations for a single sample of galaxies with homogeneously processed data, and that spans a range of masses, morphologies and SFRs.  In this way we can more fully explore the co-dependencies of the relations, and their correlations with global parameters, without the concerns that come with comparing different datasets.

To achieve our objective, we use the ALMA-MaNGA QUEnching and STar formation (ALMaQUEST) survey (Lin et al. 2020).  In Section \ref{data_sec} we review the characteristics of the ALMaQUEST survey and its galaxies, as well as providing a brief summary of the observations and data products (which are described more fully in Lin et al. 2020).  In Section \ref{ensemble_sec} we first establish the ensemble star formation relations for $\sim$ 15,000 spaxels in our sample, quantify the overall scatter, and present linear fits to these relations.  In Section \ref{pifu_sec} we investigate the rSFMS, rMGMS and rSK relations on a galaxy-by-galaxy basis (i.e. quantifying the scaling relations separately for each galaxy).  We find that all three relations show galaxy-to-galaxy variations, demonstrating that none of the couplings between \sigh2, \sigstar\ and \sigsfr\ results in a single universal law.  In Section \ref{offset_sec} we present our method for quantifying the galaxy-to-galaxy variations in the star formation scaling relations and in Section \ref{co_sec} investigate how these variations are inter-related on a galaxy-by-galaxy basis, providing insight into which relations are independent and which are due to co-variance.  Finally, in Section  \ref{stuff_sec} we investigate  how the galaxy-to-galaxy differences in the resolved star formation relations depend on the meta-properties of the host galaxy.   Our results are discussed in Section \ref{discuss_sec} and our conclusions summarized in Section \ref{summary_sec}.

\smallskip

We adopt a Salpeter initial mass function and a cosmology in which H$_0$=70 km/s/Mpc, $\Omega_M$=0.3, $\Omega_\Lambda$=0.7.

\section{Data}\label{data_sec}

The core \aq\ survey\footnote{http://arc.phys.uvic.ca/$\sim$almaquest/} consists of 46 galaxies selected from the Mapping Nearby Galaxies at Apache Point Observatory (MaNGA; Bundy et al. 2015) survey with complementary observations made by the Atacama Large Millimeter Array (ALMA) of CO (1-0).  The \aq\ galaxies are mostly in the redshift interval $0.02<z<0.05$, over which range 1 arcsecond corresponds to $\sim$ 0.5 - 1.0 kpc.  ALMA observations were designed to deliver a primary beam size (angular resolution) that is well matched to the MaNGA point spread function ($\sim$ 2.5 arcsec), allowing the ALMA data cubes to be projected onto the same spatial grids.  The sample properties, data reduction and products are described fully in the main survey paper (Lin et al., 2020), so we review only the main details here.

\subsection{ALMaQUEST Survey Sample}

\begin{figure}
	\includegraphics[width=8.5cm]{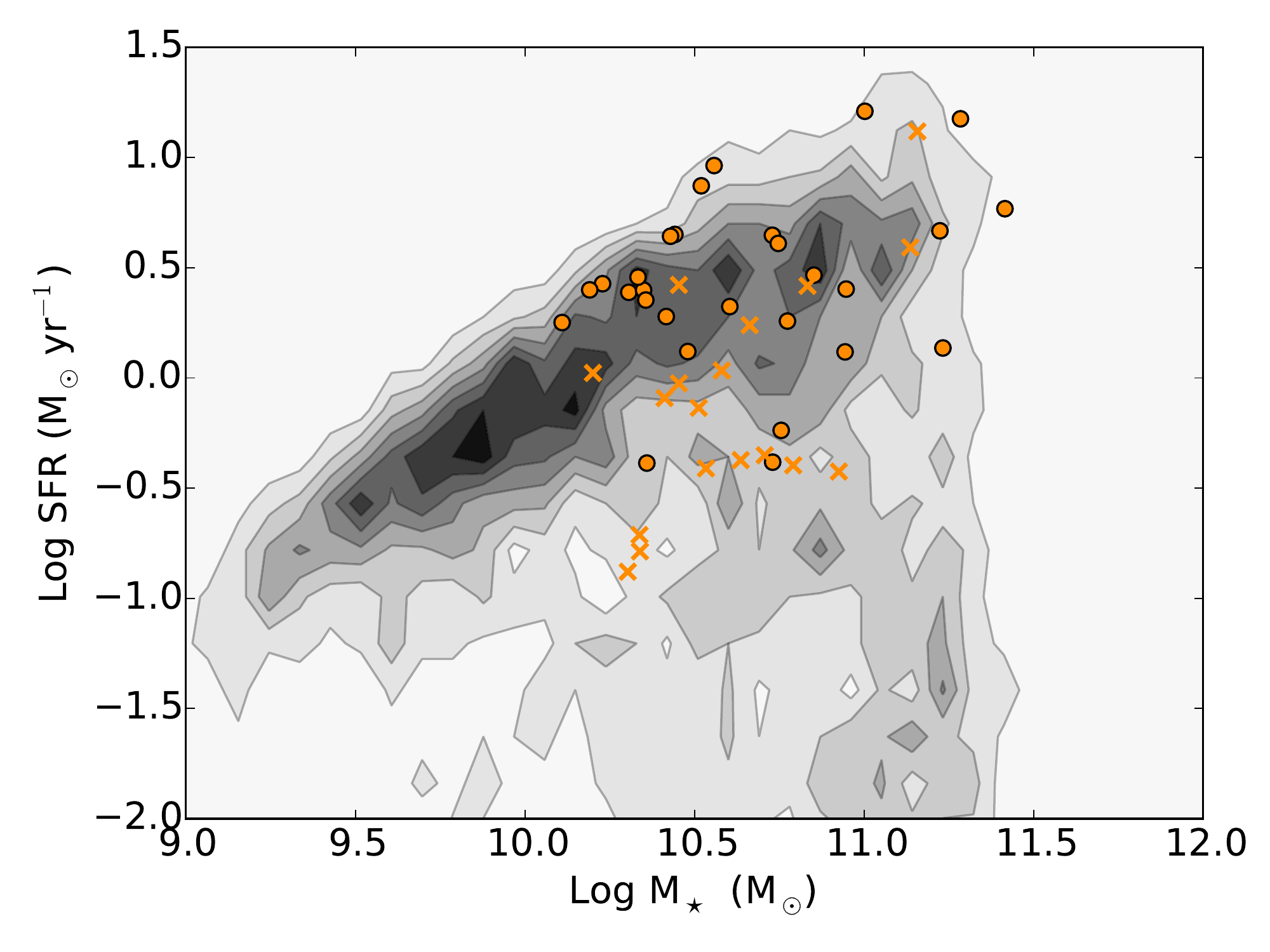}
        \caption{The global stellar mass and star formation rates for all of MaNGA DR15 (grey contours) and the \aq\ sample (symbols).  The 28 galaxies used in this study are shown as filled circles.  \aq\ sample galaxies excluded due to either high inclination or insufficent numbers of star forming spaxels are shown with crosses.}
    \label{sample_fig}
\end{figure}

The \aq\ galaxy sample was selected to include a range of specific star formation rates (sSFRs) for galaxies with total stellar masses 10 $<$ \logms\ $<$ 11.5, in order to facilitate investigation of star formation from the green valley to the starburst regime.  The targeting of galaxies both in the throes of quenching, as well as starbursts, results in a sample that is neither volumetrically complete, nor statistically representative of the galaxy population in the nearby universe.  However, the \aq\ sample is an excellent complement to more representative surveys, such as EDGE-CALIFA (Bolatto et al. 2017), which are lacking in the more extreme galaxy types.  Figure \ref{sample_fig} shows the stellar mass and star formation rates of the complete \aq\ sample (symbols) with respect to the full MaNGA DR15 sample (grey contours).  These global values of stellar mass and SFR are taken from the PIPE3D (Sanchez et al., 2016a, 2016b) value-added catalog (VAC, Sanchez et al. 2018)  and were derived through summing individual spaxel values across the MaNGA data cubes.

\subsection{MaNGA data products}

Optical emission line fluxes are taken from the public PIPE3D data cubes (S\'{a}nchez et al. 2016a,b, 2018) and corrected for internal extinction by assuming an intrinsic H$\alpha$/H$\beta$=2.85 and a Milky Way extinction curve (Cardelli, Clayton \& Mathis 1989).   Star formation rate surface densities (\sigsfr) are computed from H$\alpha$ luminosities using equation 2 from Kennicutt (1998a), a technique that has been shown to reproduce the UV and IR SFRs well in IFU data (Catalan-Torrecilla et al. 2015).   Typical errors in \sigsfr\ associated with uncertainties in the H$\alpha$ line flux are small ($<$0.02 dex), due to the large H$\alpha$ EW requirement used in our spaxel selection (see Sec 2.4).  The dominant source of uncertainty in \sigsfr\ is in the calibration itself, which may be up to several tenths of a dex (e.g. Kennicutt et al. 1998a; Brinchmann et al. 2004; Bluck et al. 2020a).  Stellar mass surface densities are taken directly from the public PIPE3D data products and have typical uncertainties of $<$ 0.07 dex, although this does not account for systematics stemming from choice of, e.g. extinction law, IMF and stellar evolution, which can be a few tenths of a dex (e.g. Mendel et al. 2014).  In order to correct for inclination effects, we compute de-projected galactocentric radii in units of both kpc and the effective radius ($R_e$), which typically extend to 1.5-2 R/R$_e$, using the Sersic fit axial ratio from the NASA Sloan Atlas (NSA).  All radii quoted in this paper adopt these de-projected values and all surface densities (e.g. \sigsfr\ and \sigstar) are inclination corrected.   

\subsection{ALMA observations and data products}

CO(1-0) spectral line observations were obtained with ALMA through programs 2015.1.01225.S, 2017.1.01093.S, 2018.1.00558.S (PI Lin) and 2018.1.00541.S (PI Ellison) between 2016-2019 in the array's second most compact configuration (C43-2).  The single pointing primary beam size for this configuration is $\sim$ 50 arcsec with an angular resolution $\sim$ 2.5 arcsec.  Integration times ranged from 0.2 to 2.5 hours on source, using one high resolution spectral window focused on the CO(1-0) line and one to three additional low resolution continuum windows for calibration.  The data cubes were all processed using the Common Astronomy Software Applications (CASA; McMullin et al. 2007) package.  The final cubes have channel widths of 11 \kms\ and root mean square (RMS) noise of $\sigma_{rms}$ = 0.2 -- 2 mJy beam$^{-1}$.  CO(1-0) luminosities were converted to molecular gas surface densities (\sigh2) via a constant conversion factor  $\alpha_{CO}$ = 4.3 M$_{\odot}$ pc$^{-2}$ (K km s$^{-1}$)$^{-1}$ (e.g. Bolatto et al. 2013) and inclination corrected in the same way as the MaNGA surface density products.  In Appendix A we re-visit the assumption of a constant conversion factor, repeating the major components of our analysis with a metallicity dependent $\alpha_{CO}$, showing that our choice of conversion factor does not significantly affect our main conclusions.

To permit a mapping of the ALMA data cubes onto the MaNGA data products, the ALMA data were first trimmed to the size of that galaxy's MaNGA cube (MaNGA IFU bundles range in size from 12 -- 32 arcsec chosen to match the galaxy size).  A fixed restoring beam size of 2.5 arcsec with pixel size of 0.5 arcsec was then applied to the ALMA cube.  These two steps resulted in ALMA data cubes with the same size and sampling as the MaNGA data products.  Typical uncertainties in \sigh2, based on the CO line flux, are $<$ 0.1 dex.  The other main source of uncertainty in \sigh2\ is in the conversion factor ($\alpha_{CO}$), the influence of which is demonstrated in the Appendix to not impact the conclusions of this work.  SDSS images and maps of data products such as \sigstar, \sigsfr\ and \sigh2, as well as further details regarding the data processing, can be found in Lin et al. (2020).

\subsection{Additional Sample Selection for This Work}

From the full \aq\ sample, we impose several additional restrictions for the work presented here.  First, we exclude highly inclined galaxies by requiring that the axial ratio listed in the NSA catalog $b/a>0.35$.  This criterion excludes 13 galaxies from the parent \aq\ sample.  Additionally, we require that any remaining galaxies have at least 20 star forming spaxels with good CO(1-0) detections, in order that we have sufficient statistics on a galaxy-by-galaxy basis to investigate resolved scaling relations in star forming regions.  To implement this criterion we first require that the S/N of the CO line intensity in a given spaxel exceeds 2.  The assessment of a spaxel as star forming proceeds in several steps\footnote{We have experimented with various versions of the star-forming spaxel criteria, such as changing the S/N cut, or the BPT threshold.  None of our results depend strongly on these choices.}.  First, we require that the spaxel has a S/N$>$2 in each of the following four optical emission lines: H$\alpha$, H$\beta$, [OIII]$\lambda$5007, [NII]$\lambda$6584.  Second, the emission line ratios for the spaxel must lie below the Kauffmann et al. (2003) designation for star formation on the classic Baldwin, Phillips \& Terlevich (1981; hereafter BPT) diagram.  Third, we impose an H$\alpha$ equivalent width (EW) cut (e.g. Cid-Fernandes et al. 2011).  We experimented with several possible cuts between 3 and 6 \AA; since 98.5\% of spaxels that pass the preceding criteria have H$\alpha$ EW $>$ 6 \AA, we adopt this more stringent cut.  We note that there are very few spaxels in our sample that are classified as AGN (according to a combination of BPT classification and requiring H$\alpha$ EW $>$ 6\AA; Cid-Fernandes et al. 2011).  A separate paper in this series (Ellison et al. 2021) investigates the rMGMS of `retired' spaxels with low H$\alpha$ EW.  Finally, we require that log \sigstar\ $>$ 7.0, a criterion that excludes a handful of anomolously small values from the PIPE3D data products.  The resulting sample consists of 28 galaxies (twice the number used to investigate scaling relations in main sequence only galaxies by Lin et al. 2019), which are shown as filled circles in Fig. \ref{sample_fig}.  Full target details and ALMA data details are given in Lin et al. (2020), but for convenience we summarize the main properties in Table \ref{target_table}, including the number of spaxels used in each galaxy in this work. Combined, these 28 galaxies contain 15,035 spaxels with measured values of \sigh2, \sigstar\ and \sigsfr.  Galaxies from the \aq\ survey that are/are not used in the current work are shown as circles/crosses in Fig \ref{sample_fig}.  

\begin{table}
  \caption{Summary of global properties for ALMaQUEST galaxies used in this work. The number of spaxels indicates the number in each galaxy that pass the various detection and S/N thresholds required in both CO and in optical emission lines.}
\begin{tabular}{lccccr}
\hline
Plate-ifu  &  $z$  & log(M$_{\star}$/M$_{\odot}$) & log (SFR/yr) &  Sersic N$_s$ & \# spaxels\\
\hline
8241-3703  &  0.02911 &  10.11 &   0.25 &  1.5  & 555  \\
8615-3703  &  0.01845 &  10.19 &   0.40 &  1.6  & 291  \\
8084-3702  &  0.02206 &  10.23 &   0.43 &  1.5  & 167  \\
8082-6103  &  0.02416 &  10.31 &   0.39 &  1.4  & 784  \\
8952-6104  &  0.02843 &  10.33 &   0.46 &  2.2  & 713  \\
7977-3703  &  0.02782 &  10.35 &   0.40 &  0.9  & 653  \\
8155-6102  &  0.03081 &  10.36 &   0.35 &  1.7  & 1041  \\
7977-3704  &  0.02724 &  10.36 &  -0.39 &  3.3  & 305  \\
8655-9102  &  0.04505 &  10.42 &   0.28 &  0.8  & 330  \\
8450-6102  &  0.04200 &  10.43 &   0.64 &  1.0  & 633  \\
8616-9102  &  0.03039 &  10.44 &   0.65 &  3.2  & 823  \\
8082-12701 &  0.02703 &  10.48 &   0.12 &  3.01 & 1038  \\
8156-3701  &  0.05273 &  10.52 &   0.87 &  0.9  & 424  \\
8081-3704  &  0.05400 &  10.56 &   0.96 &  2.0  & 222  \\
8081-9101  &  0.02846 &  10.60 &   0.32 &  1.8  & 269  \\
8077-6104  &  0.04601 &  10.73 &   0.65 &  1.3  & 876  \\
8952-12701 &  0.02856 &  10.73 &  -0.38 &  1.61 & 63  \\
8078-6103  &  0.02859 &  10.75 &   0.61 &  1.9  & 932  \\
8616-12702 &  0.03083 &  10.76 &  -0.24 &  4.42 & 190  \\
8616-6104  &  0.05426 &  10.77 &   0.26 &  0.8  & 348  \\
7977-12705 &  0.02724 &  10.85 &   0.47 &  4.85 & 367  \\
8086-9101  &  0.04003 &  10.94 &   0.12 &  6.0  & 375  \\
8078-12701 &  0.02698 &  10.95 &   0.40 &  3.51 & 705  \\
8241-3704  &  0.06617 &  11.00 &   1.21 &  1.9  & 561  \\
8083-12702 &  0.02104 &  11.22 &   0.67 &  2.62 & 1893  \\
7977-9101  &  0.02656 &  11.23 &   0.14 &  3.8  & 165  \\
8623-6104  &  0.09704 &  11.28 &   1.18 &  3.7  & 276  \\
8082-12704 &  0.13214 &  11.42 &   0.77 &  5.34 & 36  \\
\hline
\end{tabular}
\label{target_table}
\end{table}

\section{Ensemble resolved star formation scaling relations}\label{ensemble_sec}

\begin{figure*}
	\includegraphics[width=5.5cm]{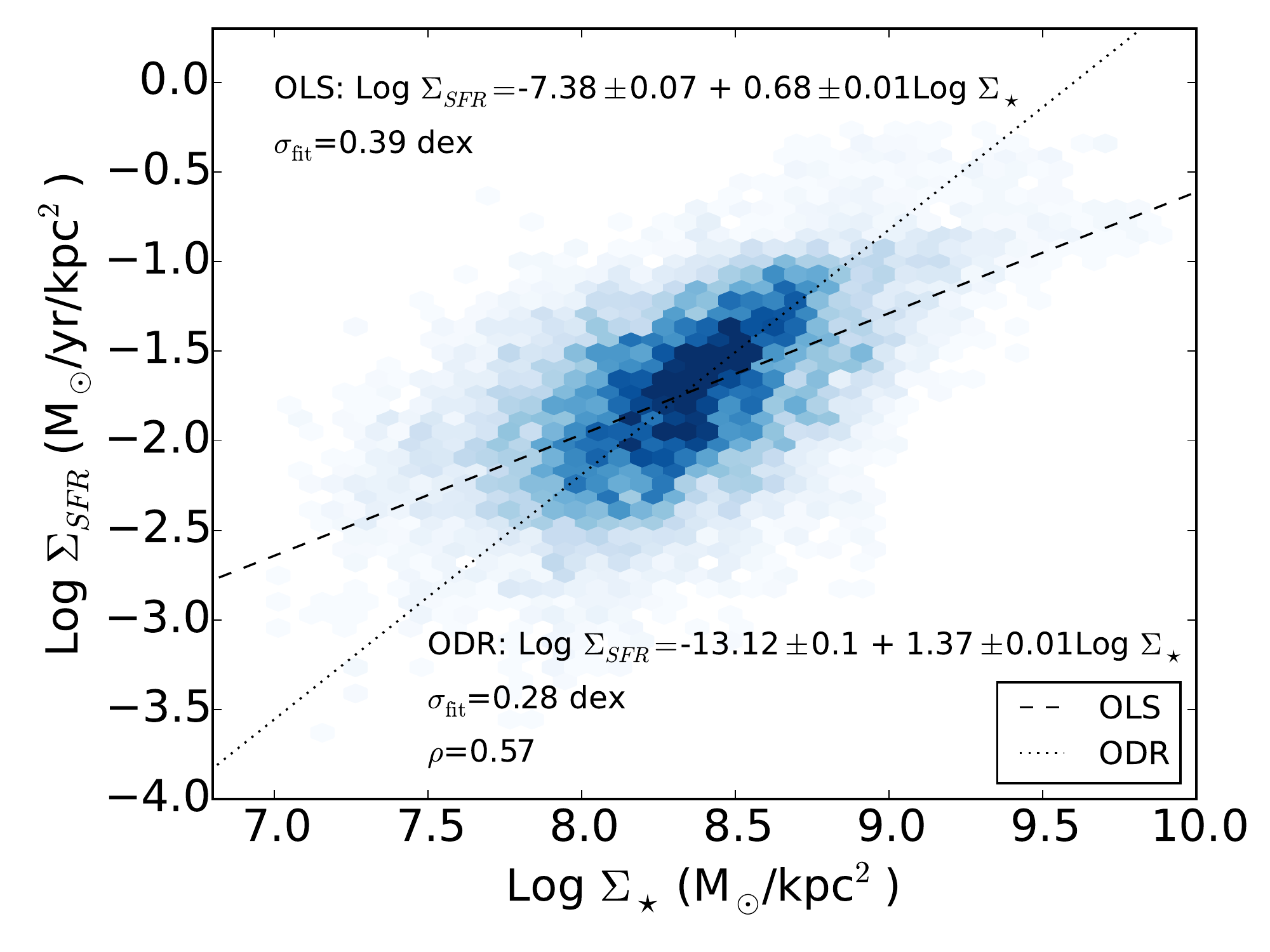}
	\includegraphics[width=5.5cm]{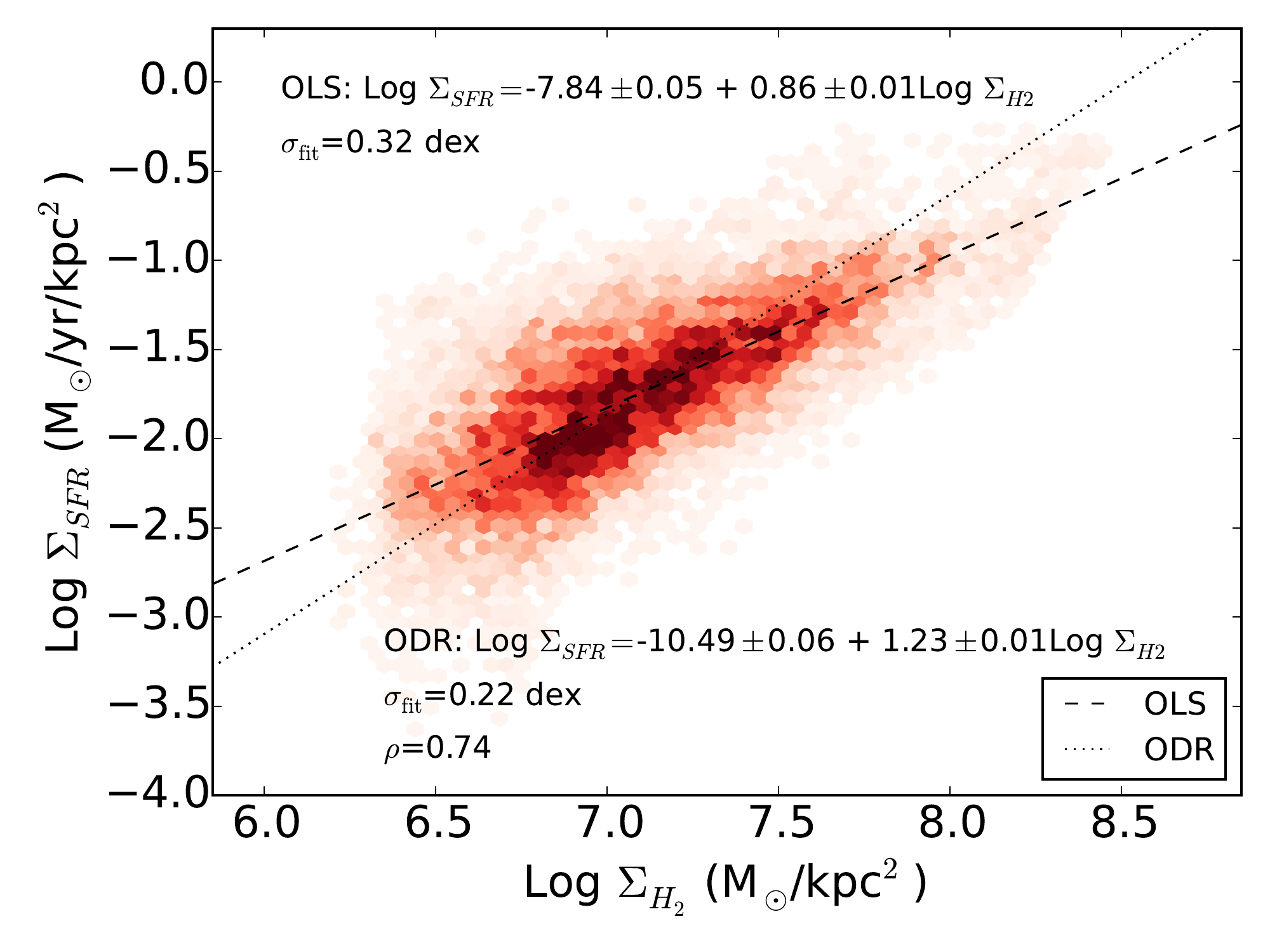}
	\includegraphics[width=5.5cm]{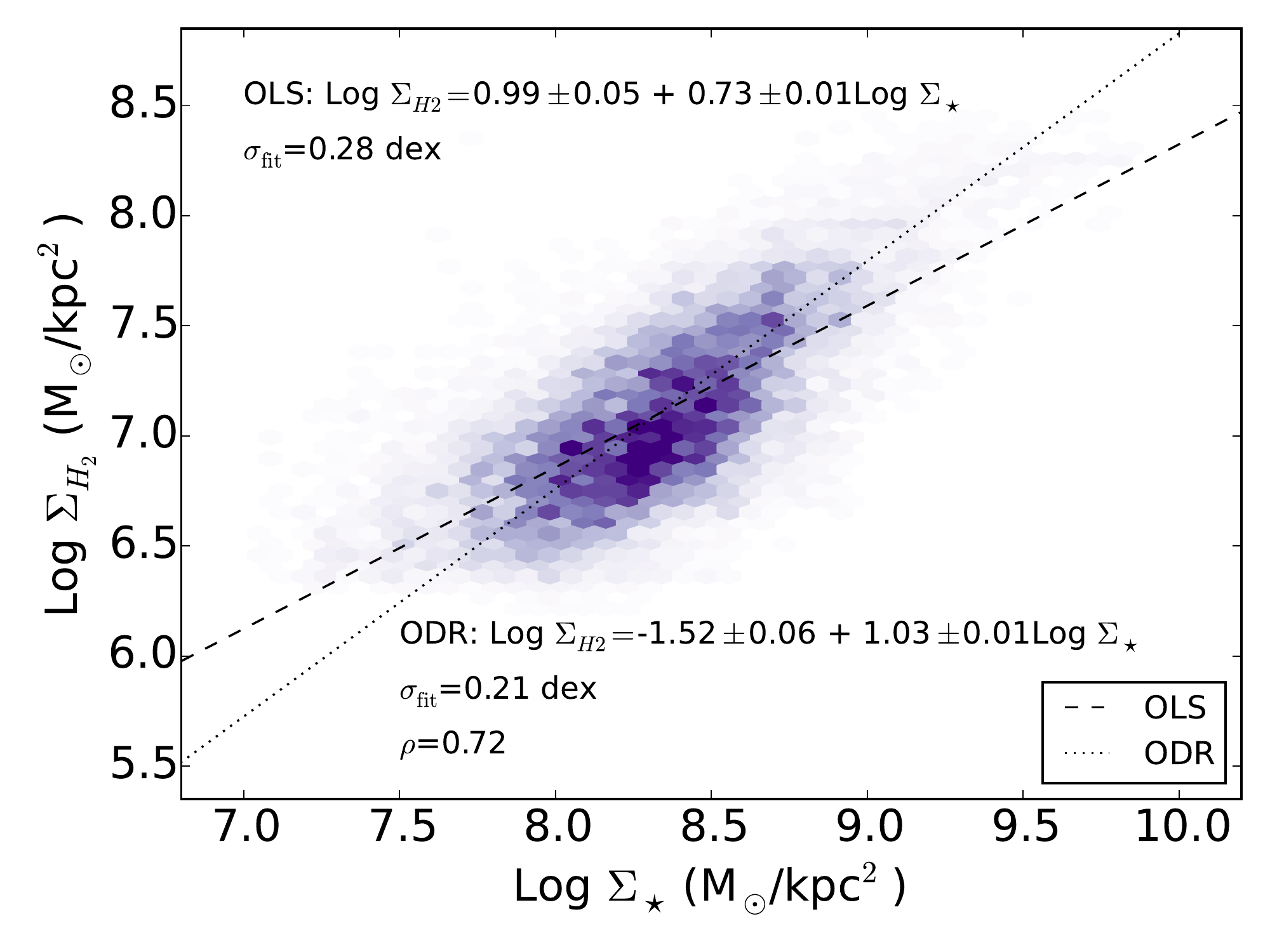}
        \caption{Ensemble star formation relations for 15,035 spaxels in our sample of \aq\ 28 galaxies.  Left panel: The resolved star-forming main sequence.  Middle panel: the resolved Schmidt-Kennicutt relation.  Right panel: The resolved molecular gas main sequence. The ordinary least squares and the orthogonal distance regression fits are shown by dashed and dotted lines respectively.  The coefficients (and their errors) of the OLS and ODR fits are reported in the top and bottom each panel, along with the scatter around each best fit ($\sigma_{fit}$), as well as the Pearson correlation coefficient ($\rho$). }
    \label{ensemble_fig}
\end{figure*}

Although the main goal of this paper is to study the resolved star formation relations for individual galaxies, it is useful to first establish a reference for the ensemble of all spaxels used in this work.  However, it is important to preface the presentation of our spaxel ensembles with several caveats that affect the star formation scaling relations in our sample.

First, since resolved relations depend on global galaxy parameters (e.g. Gonzalez Delgado et al. 2016; Bolatto et al. 2017; Cano Diaz et al. 2019), the resolved relations presented for the ALMaQUEST sample will therefore likely differ from those derived for large representative samples.  Second, we have specifically chosen our spaxel sample to include only data for which \sigstar, \sigsfr\ and \sigh2\ could all be reliably measured, and for which the ionization source was deemed to be star formation.  Spaxels that are ionized either by AGN (e.g. Sanchez et al. 2018) or by old stellar populations (e.g. Hsieh et al. 2017; Cano-Diaz et al. 2019; Ellison et al. 2021), are known to be offset from the locus of star forming regions.

Finally, a slew of technical details, such as the fitting method and its functional form (e.g. Hsieh et al. 2017; Erroz-Ferrer et al. 2019), weighting/binning schemes, variable IFU coverage, S/N cuts, detection thresholds in the data and choice of variable range for fitting (e.g. Cano Diaz et al. 2019) can all affect the derived coefficients to the resolved scaling relations.  These effects have been extensively explored in detail in the works cited above.  In contrast, the main goal of the current paper is to investigate the galaxy-to-galaxy variation of the resolved scaling relations, rather than to present a detailed re-assessment of the data ensemble.  We therefore do not explore these numerous selection effects in our own data set and present the fits to our ensemble data sample for reference only.

\subsection{The ensemble resolved star formation main sequence}

In Fig. \ref{ensemble_fig} we present the rSFMS (left panel), rSK relation (middle panel) and rMGMS (right panel) for the 15,035 spaxels in our sample.  We will use the same colour-coding (rSFMS: blue, rSK relation: red and rMGMS: purple) in figures throughout this paper for consistency.  The ensemble relations are fit using two different methods: an ordinary least squares (OLS) approach and an orthogonal distance regression (ODR) fit.  The former of these methods assumes that the $x$-variable is known to infinite precision, i.e. it has no error.  The fit is therefore optimized to minimize scatter in the $y$-variable.  In contrast, the ODR assumes that there can be uncertainties in both the $x$- and $y$-variables and hence minimizes the scatter in both variables.  Despite the $x$-$y$ plotting conventions for the three star formation relations, both quantities have uncertainties and the relations could equally well be plotted with flipped axes (see Shetty et al. 2013 for further discussion on fitting methods in the context of the rSK relation).  An ODR fit is therefore sometimes preferred (e.g. Lin et al. 2019; Morselli et al. 2020), and can yield quite different results to an OLS approach (e.g. Hsieh et al. 2017).

In Fig. \ref{ensemble_fig} we present the results (coefficients and scatter) of the OLS (dashed line) and ODR (dotted line) fits for all three star formation scaling relations.  Visually, the ODR fits are a better representation of the data (as previously found for the rSFMS by Hsieh et al. 2017).  The ODR fits have a steeper slope than the OLS fits for all three relations, as well as a smaller scatter.  Cano-Diaz et al. (2019) have recently suggested that an artificial flattening of the rSFMS in MaNGA data can be introduced by including spaxels with log \sigstar\ $<$ 7.5 (and in general, a curvature at the low end of any relation can be introduced by datasets of different depths).  We recall that our sample already requires log \sigstar\ $>$ 7.0, such that only $\sim$ 500 spaxels (out of $\sim$15,000) are excluded by imposing this more stringent mass cut.  Repeating our fits with this stricter cut of log \sigstar\ $>$ 7.5 has a minimal effect on the OLS slope, changing it from 0.68 to 0.70.  However, the ODR fit is more impacted, and exclusion of log \sigstar\ $<$ 7.5 spaxels steepens the slope of the fit from 1.37 to 1.50.

There are numerous versions of the rSFMS already present in the literature (e.g. Sanchez et al. 2013; Cano-Diaz et al. 2016, 2019; Gonzalez-Delgado et al. 2016; Hsieh et al. 2017; Erroz-Ferrer et al. 2019; Lin et al. 2019) with slopes that vary from $\sim$0.6 to $\sim$1.3.  Variation between the fits presented in these works likely has contributions from different sample properties, selection effects and fitting methods, which can affect the derivation of fit coefficients (as demonstrated by our comparison of the OLS and ODR methods in Fig. \ref{ensemble_fig}).  Moreover, Hani et al. (2020) have used simulations to demonstrate that the slope of the rSFMS further depends on both the SFR indicator used and on the spatial resolution of the sampling.    Likewise, the scatter around the fit is particularly sensitive to sample selection (as we will show later in this paper, the rSFMS depends on parameters such as total stellar mass and morphology) and the details of the selection of star-forming spaxels.

With these caveats in mind, we refrain from a detailed comparison of the ensemble rSFMS of our sample with previous works, and simply note that our derived slope ($\sim$0.7, 1.4 for the OLS and ODR respectively) and scatter ($\sim$0.4, 0.3 dex for the OLS and ODR respectively) are within the range found by previous works.  The slope derived from the ODR method is in reasonable agreement with the value ($\sim$1.2) derived by Lin et al. (2019) for the subset of $\sim$ 5400 star-forming spaxels in 14 \aq\ main sequence galaxies.  The ODR fit of the rSFMS corresponds to a specific SFR (sSFR = \sigsfr/\sigstar) of $-10.0$ at log \sigstar\ = 8.5.

\subsection{The ensemble resolved Schmidt-Kennicutt relation}

The slope of the rSK relation has been the focus of intense disussion for many years, due to its implication for a universal (or not) star formation recipe.   A rSK relation slope of one, both within a given galaxy and between different galaxies, would indicate that the conversion between gas and stars can be characterized by a single efficiency (or depletion time, where $\tau_{\rm dep}$ = \sigh2/\sigsfr\ = 1/SFE).  Conversely, a slope steeper than unity implies that higher surface densities of gas lead to even higher surface densities of star formation, indicating that star formation is enhanced by non-linear processes that operate in regions of high gas density.  It is also often pointed out that a simple dynamical model in which stars form with a characteristic timescale that is the freefall time, and under the assumption of a uniform scale height, leads to the prediction of a slope of 1.5 (although other theoretical assumptions can yield different slopes, e.g. Tan 2000). 

Observationally, some of the earliest measurements of the slope of the \textit{global} SK relation from total gas (HI and CO) observations yielded a slope of 1.3 -- 1.4, tantalizingly close to the simple theoretical expectation of an N=1.5 slope (e.g. Kennicutt 1989, 1998a,b).  A super-linear slope has also been reported in some works on the \textit{resolved} SK relation (e.g. Momose et al. 2013).  A modern re-assessment of total gas surface densities has confirmed this relatively steep slope (de los Reyes \& Kennicutt 2019).  However, de los Reyes \& Kennicutt (2019) also find that using molecular gas alone yields a significantly flatter SK relation, with a slope closer to unity, although the inclusion of starbursting galaxies can drive a much steeper relation (e.g. Daddi et al. 2010).  Likewise, studies of the rSK relation for ensembles of nearby galaxies have generally determined a rSK relation between \sigh2\ and \sigsfr\ with slope $\sim$1, indicating a universal molecular gas star formation efficiency with a characteristic depletion time of 1-2 Gyr in ensemble datasets (e.g. Bigiel et al. 2008, 2011; Schruba et al. 2011; Leroy et al. 2013).   However, as with the rSFMS, it has been shown that the slope of the rSK relation is also sensitive to the resolution of the data (Calzetti, Liu \& Koda 2012).

In the middle panel of Fig. \ref{ensemble_fig} we present the rSK relation for the 15,035 star forming spaxels in our sample of 28 \aq\ galaxies.  Once again we note that the ODR fit yields a steeper relation (slope$\sim$1.2) compared with the OLS fit (slope$\sim$0.9).  The ODR fit leads to a depletion time of 0.7 Gyr for log \sigh2=7.0.  Although the rSK relation determined herein is in reasonable agreement with previous works (e.g. Bigiel et al. 2008, 2011; Schruba et al. 2011; Leroy et al. 2009, 2013; Usero et al. 2015; Lin et al. 2019; Morselli et al. 2020), we once again caution that the details of the fitting procedure can significantly influence the derived coefficients  (e.g. Leroy et al. 2013; de los Reyes \& Kennicutt 2019).

\subsection{The ensemble resolved molecular gas main sequence}

Few studies have explored the rMGMS.  Although first demonstrated by Shi et al. (2011) and Wong et al. (2013), the first parametrization of a large ensemble of data was presented by Lin et al. (2019) for $\sim$ 5400 spaxels in 14 main sequence galaxies in the \aq\ sample, and soon after by Morselli et al. (2020) for 5 nearby disk galaxies.  These works derived similar slopes for the rMGMS of 1.1 and 0.9 respectively (both using an ODR fit), with a scatter of about 0.2 dex in each case.  The scatter around the rMGMS found by Lin et al. (2019) and Morselli et al. (2020) is marginally higher than the scatter around the rSK relation in the same samples (by a few hundredths of a dex).

\begin{figure*}
	\includegraphics[width=19cm]{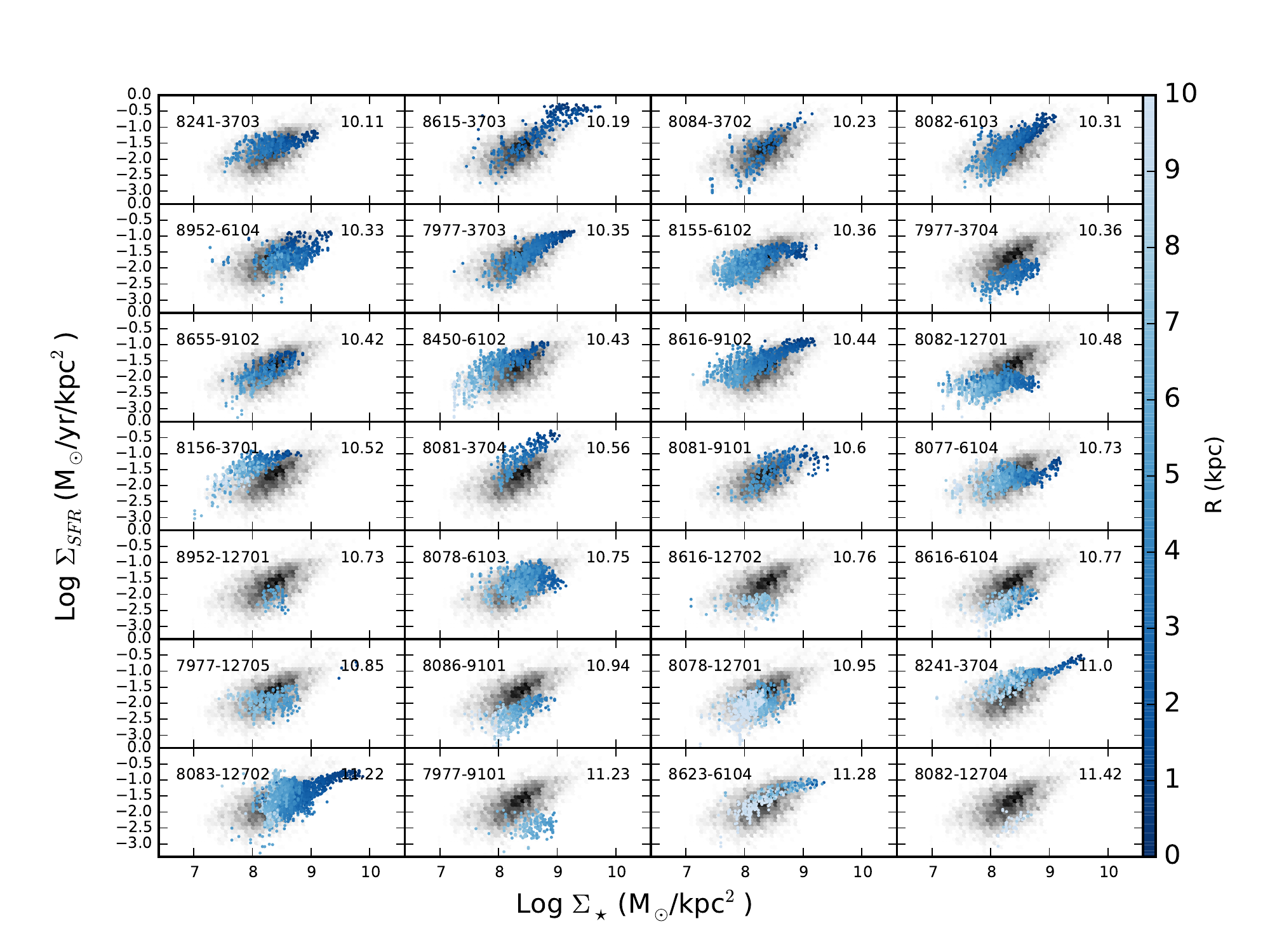}
        \caption{The resolved star forming main sequence for each of the 28 galaxies in our sample.  The MaNGA plate-IFU identifier is given in the top left of each panel.  The panels are ordered by increasing total stellar mass (from the PIPE3D VAC), which is given in the top right of each panel.   The background greyscale shows the number density in the combined sample of $\sim$ 15,000 spaxels and is used for visual reference in each panel.  The blue points show the individual galaxy relations and shading indicates the galactocentric radius in units of kpc.  There is a large galaxy-to-galaxy diversity in the rSFMS (both shape and normalization), even at approximately fixed stellar mass.}
    \label{rSFMS_fig}
\end{figure*}

In the right panel of Fig. \ref{ensemble_fig} we present the rMGMS for the $\sim$15,000 spaxels in our \aq\ sample, which contains the largest number of galaxies used to date to explore this relation.  Once again, we find the ODR yields a better visual fit to the data and a slightly steeper slope, whose value (1.0) is in good agreement with Lin et al. (2019) and Morselli et al. (2020).   The ODR fit to the rMGMS corresponds to a molecular gas-to-stellar mass fraction (hereafter, simply `molecular gas fraction', \fgas\ = \sigh2/\sigstar) of 0.06 at log \sigstar\ = 8.5.

The scatter of the rMGMS around either the ODR or the OLS fit is smaller than the scatter for the corresponding method's fits to the other two relations.    However, for the ODR fit, the difference in the scatter between the rMGMS and rSK relations is minimal: 0.21 dex vs. 0.22 dex.   These values are similar to those reported by Lin et al. (2019) and Morselli et al. (2020): 0.20 and 0.22 dex, respectively, where the Lin et al. (2019) sample uses the main sequence subset of \aq\ galaxies.  We therefore agree with the previous conclusion of Lin et al. (2019) that the rSK relation and rMGMS have very similar (small) scatters, compared to larger deviations seen around the rSFMS.   A proposed interpretation of these comparative scatters is that the rSFMS is a manifestation of the more fundamental rSK relation and rMGMS (Lin et al. 2019; Morselli et al. 2020).  Also in support of this interpretation are the Pearson correlation coefficients which are found to be stronger for the rMGMS (0.72) and rSK (0.74) relation than for the rSFMS (0.57).  We return in Section \ref{co_sec} with further complementary evidence that the rSK and rMGMS represent independent physical mechanisms, whereas the rSFMS is the result of covariance.

\section{Galaxy-by-galaxy resolved scaling relations}\label{pifu_sec}

\begin{figure*}
	\includegraphics[width=19cm]{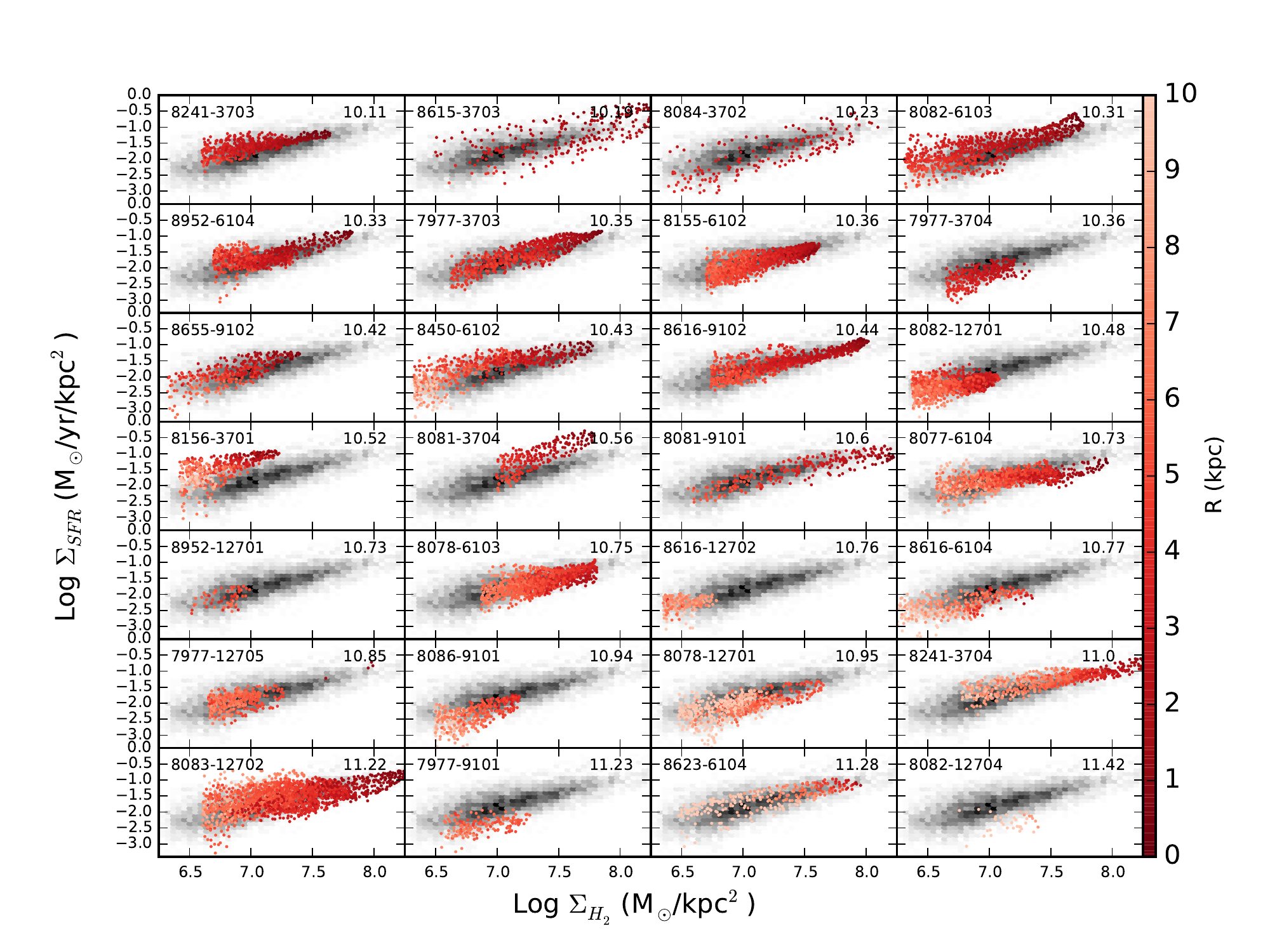}
        \caption{The resolved Schmidt-Kennicutt relation for each of the 28 galaxies in our sample.  The MaNGA plate-IFU identifier is given in the top left of each panel.  The panels are ordered by increasing total stellar mass (from the PIPE3D VAC), which is given in the top right of each panel.    The background greyscale shows the number density in the combined sample of $\sim$ 15,000 spaxels and is used for visual reference in each panel.  The red points show the individual galaxy relations and shading indicates the galactocentric radius in units of kpc. There is a large galaxy-to-galaxy diversity in the rSK relation, even at approximately fixed stellar mass.}
    \label{rKS_fig}
\end{figure*}

\begin{figure*}
	\includegraphics[width=19cm]{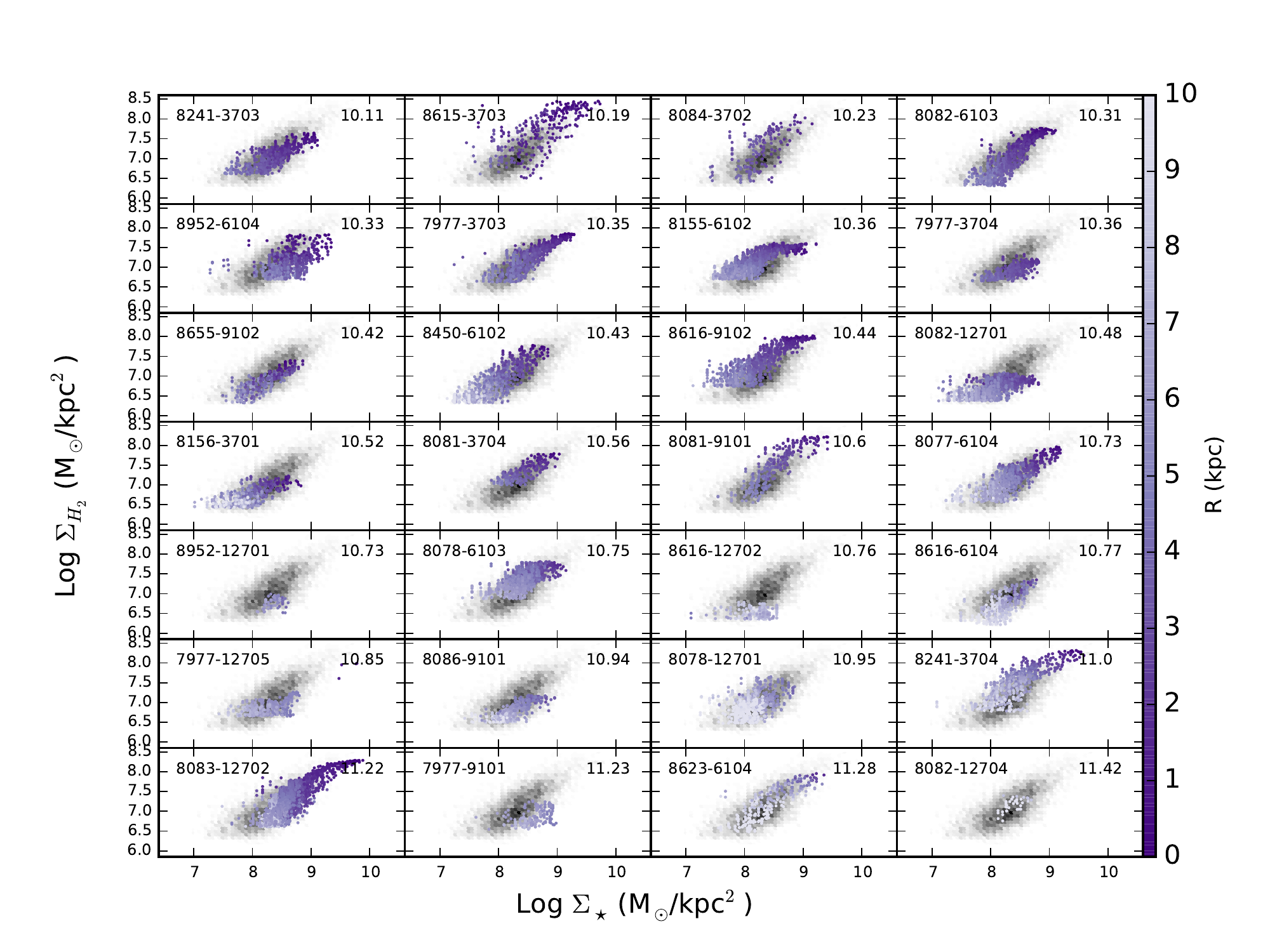}
        \caption{The resolved molecular gas main sequence for each of the 28 galaxies in our sample.  The MaNGA plate-IFU identifier is given in the top left of each panel.  The panels are ordered by increasing total stellar mass (from the PIPE3D VAC), which is given in the top right of each panel.    The background greyscale shows the number density in the combined sample of $\sim$ 15,000 spaxels and is used for visual reference in each panel.  The purple points show the individual galaxy relations and shading indicates the galactocentric radius in units of kpc. There is a large galaxy-to-galaxy diversity in the rMGMS relation, even at approximately fixed stellar mass.}
    \label{rMGMS_fig}
\end{figure*}

Having established the overall properties of the ensemble of our sample, we now turn to an inspection of the three star formation relations on a galaxy-by-galaxy basis.   Comparison of these scaling relations between galaxies will enable an assessment of whether a universal set of `recipes' exists for the distribution of stellar mass, molecular gas and the formation of new stars.

\subsection{The resolved star forming main sequence}

Fig. \ref{rSFMS_fig} shows the rSFMS for each of the 28 galaxies in our sample, with the MaNGA plate-IFU identifier given in the top left of each panel.   The panels are ordered by increasing total stellar mass, whose value is indicated in the top right of each panel.  In each panel the greyscale background shows the distribution of \sigstar\ and \sigsfr\ for the full sample of 15,035 spaxels for reference (i.e. identical to the left panel of Fig. \ref{ensemble_fig}). The blue points in each panel show the spaxels for each individual galaxy, with the shade indicating the (de-projected) distance from the galaxy centre in kpc (the figure looks qualitatively similar if points are instead colour coded by effective radius).  As expected, higher values of \sigstar\ are seen at smaller galactocentric radii.  Although the individual spaxels are presented in Fig. \ref{rSFMS_fig}, it is important to emphasize that, since the spaxel scale of our data products (0.5 arcsec) somewhat over-samples the resolution of the data (FWHM $\sim$ 2-2.5 arcsec), and that these data points are therefore not independent.  In the quantitative analyses presented later in this paper, the data are smoothed, rather than considering individual spaxels. Experiments with coarser binning the data lead to consistent results.

It is immediately obvious that, although a rSFMS exists for almost all of the galaxies in our sample, there is significant galaxy-to-galaxy variation, even at approximately fixed total stellar mass.  For example, the galaxies in the 3rd row from the top in Fig. \ref{rSFMS_fig} span a total stellar mass of only 0.06 dex (10.42 $<$ \logms\ $<$ 10.48), and yet the rSFMS of the four galaxies plotted in this row vary from being normal (1st column), enhanced (2nd and 3rd columns), or suppressed (4th column) relative to the reference distribution shown in grey.  In general, the galaxy-to-galaxy variation affects the entire population of star forming spaxels in a given galaxy (although the radii of these spaxels can vary), changing the overall normalization of the relation for all the spaxels.

In addition to the variation in the normalization of the galaxy-to-galaxy rSFMS, the shapes also exhibit significant diversity.  Several of the galaxies in our sample show a saturation, or flattening, of \sigsfr\ at high \sigstar.  However, other galaxies exhibit a much more linear relation, with \sigsfr\ values that exceed the `saturation' point seen in other galaxies. For example, 8616-9102, 7977-3703 and 8623-6104 all have star formation rate surface densities that saturate at log \sigsfr $\sim -1$ $M_{\odot}$ yr$^{-1}$ kpc$^{-2}$.  In contrast 8615-3703, 8082-6103 and 8081-3704 all have star formation rate surface densities that continue to rise linearly with \sigstar\ beyond this ceiling.  Whether or not a galaxy shows a turnover in its \sigsfr\ at high \sigstar\ seems to be independent of its mass or Sersic index (Table \ref{target_table}).  

Visual inspection of the SDSS images of the galaxies in our sample (see Lin et al. 2020 for the full data presentation) reveals that most fall into the expected colour-morphology paradigm, with disks that are blue and actively star forming and bulges that have a redder colour.  Such galaxies understandably exhibit the flattened rSFMS relations exhibited by, for example 8616-9102, 7977-3703 and 8623-6104, where the central high \sigstar\ regions have reduced sSFR.  The role of a low sSFR bulge has previously been suggested to contribute to a flattening of the global SFMS at high stellar mass (Abramson et al. 2014; Pan et al. 2018; Leslie et al. 2020).  However, some galaxies in our sample have distinctly blue central regions, including 8615-3703, 8082-6103 and 8081-3704 (the former of these is clearly a late stage merger).  In these galaxies, star formation is apparently continuing rigourously even in the central regions, leading to high \sigsfr\ even at high \sigstar.  We return briefly to the diversity of radial properties in Section \ref{rMGMS_sec}, although defer a full examination of radial properties to a forthcoming paper.  In summary, Fig. \ref{rSFMS_fig} shows that although there is clearly a coherent rSFMS amongst star-forming spaxels within each galaxy, neither the shape, nor the normalization is universal.  

In addition to the galaxy-to-galaxy variation of the rSFMS, there is evidently considerable vertical scatter around the relation within a given galaxy.  In Lin et al. (2020) we present the distributions of sSFR for each of the galaxies in our sample.  Lin et al. (2020) show that there is frequently at least an order of magnitude spread in sSFR within a given galaxy.  Although part of this diversity in sSFR within a given galaxy is due to radial differences, as evidenced by the flattening of the rSFMS at high \sigstar\ that is seen in many of the galaxies in Fig. \ref{rSFMS_fig}, we can also discern that there can be a broad range of sSFR even at a fixed \sigstar\ (or radius).

\subsection{The resolved Schmidt-Kennicutt relation}

Fig. \ref{rKS_fig} is analogous to  Fig. \ref{rSFMS_fig}, but shows the rSK relation for each of the 28 galaxies in our sample.  The greyscale background is again the combined distribution of all $\sim$ 15,000 spaxels across all 28 galaxies, shown in all panels for reference.  The ordering of galaxies is the same as in Fig. \ref{rSFMS_fig}.   As for the rSFMS, there is again significant galaxy-to-galaxy variation, although visually the variation is less dramatic than for the rSFMS (we will return to quantify this point in the next Section).  Fig. \ref{rKS_fig} confirms numerous previous works that have shown that the rSK is not universal (or at least requires more parameters to capture its diversity, e.g. Dey et al. 2019).

Fig. \ref{rKS_fig} reveals that there is some variation in the depths to which the molecular gas is detected, with effective molecular gas surface density thresholds that range from log \sigh2\ $\sim$ 6.3 to 6.8 M$_{\odot}$/kpc$^2$ on a galaxy by galaxy basis.  Despite this variation it is clear from Fig. \ref{rKS_fig} that there are systematic variations in the `recipe' that galaxies follow in converting their molecular gas into stars, at least as traced by CO(1-0), and that these differences operate over all radii.  

In the previous subsection, we showed that \sigsfr\ saturates at high \sigstar\ for many galaxies in our sample.  However, the rSK relation shows no such \sigsfr\ saturation at high \sigh2.  Even galaxies that exhibit a flattening of their rSFMS, such as 8616-9102, 7977-3703 and 8623-6104, have smoothly rising rSK relations.  Therefore, unlike the rSFMS and (as we will show in the next subsection) the rMGMS, the shape of the rSK relation for the galaxies in our sample is relatively invariant, although its slope and normalization change from galaxy-to-galaxy (as we will quantify in the next Section).

Fig. \ref{rKS_fig} shows that there is significant vertical scatter around the rSK relation, indicating a variable SFE even within a given galaxy.  The full distribution of SFEs within a given galaxy is presented in Lin et al. (2020), where it is shown that SFEs can vary by an order of magnitude.  Two galaxies in our sample show a particularly large scatter in their rSK relations in Fig. \ref{rKS_fig}: 8615-3703 and 8084-3702 (top row, 2nd and 3rd columns).  Both of these galaxies show signs of recent merger activity with strong central dust lanes (see galaxy images in Lin et al. 2020).  8615-3703 was also identified by Ellison et al. (2020a) as hosting a central starburst.   

\subsection{The resolved molecular gas main sequence}\label{rMGMS_sec}

To complete the trifecta of resolved scaling relations studied in this paper, in Fig. \ref{rMGMS_fig} we present the rMGMS for the 28 galaxies in our sample.  As in previous figures, the galaxies are ordered by total stellar mass and we display all $\sim$15,000 spaxels in the sample as a grey scale in all of the panels as a reference (i.e. the same distribution as shown in the right panel of Fig. \ref{ensemble_fig}).  Individual galaxy relations are shown by purple points, with shading to indicate their galactocentric distance in units of kpc.  Variation between galaxies is again evident, although visually there appears to be less scatter than in the rSFMS or rSK relations presented in the preceding sub-sections.  We will return quantitatively to this point in the next Section.  Again we note that in addition to the galaxy-to-galaxy variation in the rMGMS, there is significant vertical scatter in \sigh2\ at fixed \sigstar, indicating that a broad range of molecular gas fractions are present.  The full distribution of \fgas\ for each galaxy is presented in Lin et al. (2020).

\begin{figure}
  \includegraphics[trim=0cm 4cm 0cm 5cm,clip=true,width=9cm]{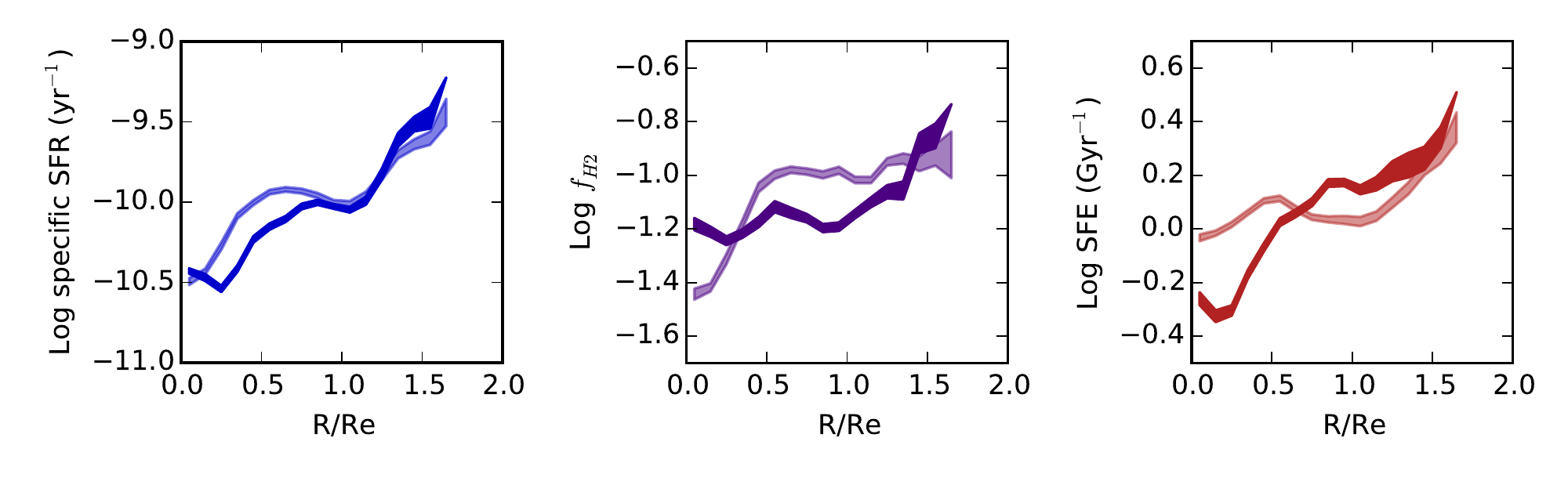}
        \caption{Radial profiles of sSFR, \fgas\ and SFE for the spaxels of two galaxies in our sample: 8155-6102 (pale curves) and 8077-6104 (dark curves), chosen to demonstrate contrasting behaviours in \fgas\ and SFE. Both galaxies exhibit a suppressed sSFR in their central regions (left panel).  However, the reason for this suppression appears to be different for the two galaxies: a low central gas fraction for 8155-6102, but a low central SFE for 8077-6104.} 
    \label{profiles_fig}
\end{figure}

As noted for the rSFMS, the shape of the rMGMS deviates from a single gradient for many of the galaxies and exhibits a flattening at high \sigstar.  A similar saturation of \sigh2\ at high \sigstar\ was also seen in the rMGMS for individual galaxies by Wong et al. (2013).  As discussed in the previous sub-section, no such flattening is seen in the rSK relations, with \sigsfr\ continuing to increase at the highest \sigh2\ values. Taken together, these results might initially be taken to suggest that the reduced sSFRs seen at high \sigstar\ (i.e. small galactocentric radius) are therefore primarily due to a reduction in gas fraction.  However, closer inspection indicates that central sSFRs may be lowered either by low gas fractions or by lower SFEs.   In Fig. \ref{profiles_fig} we show the sSFR, \fgas\ and SFE profiles of two galaxies in our sample, 8155-6102 and 8077-6104, to illustrate this point.  Profiles are plotted as a function of (de-projected) effective radius for a clearer comparison.  Both galaxies show a suppresion in their central sSFRs (left panels).  However, the reason for this suppression appears to differ for the two galaxies.  In 8155-6102 (pale curves of Fig. \ref{profiles_fig}) the profile of SFE is fairly flat, but the gas fraction drops in the centre.  Conversely, in 8077-6104 (dark curves of Fig. \ref{profiles_fig}) the gas fraction is fairly constant within the effective radius, but the SFE is suppressed in the centre.  A more extensive study of the radial profiles of star forming and green valley galaxies is presented in Pan et al. (in prep).  For now, we simply highlight that the diversity in galaxy-to-galaxy star formation scaling relations, and their radial properties, are likely driven by a variety of processes (see also Leroy et al. 2013; Utomo et al. 2017; Colombo et al. 2018).  

\section{Quantifying galaxy-to-galaxy variations for the resolved star formation relations}\label{offset_sec}

\begin{figure*}
	\includegraphics[width=5.5cm]{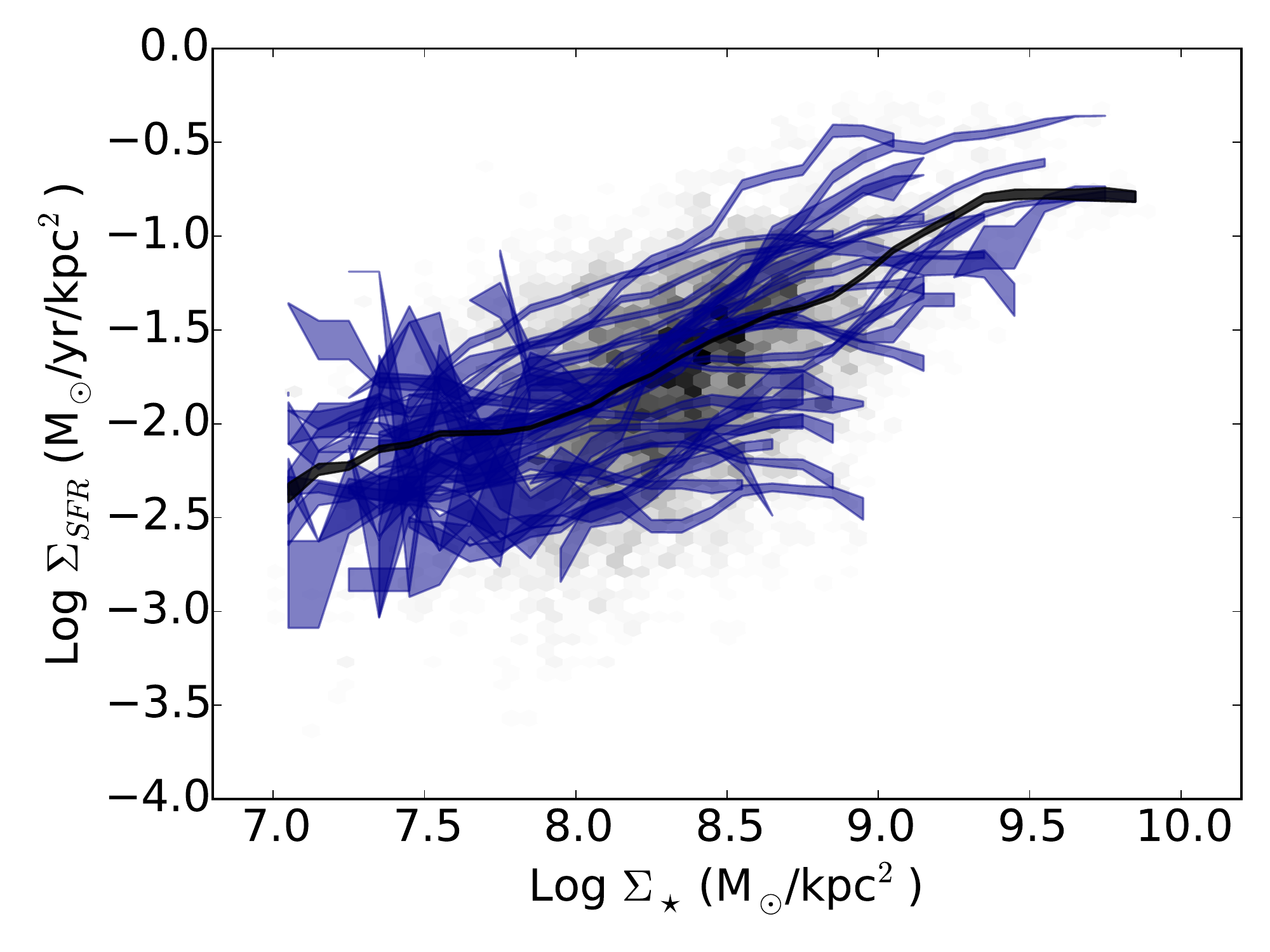}
	\includegraphics[width=5.5cm]{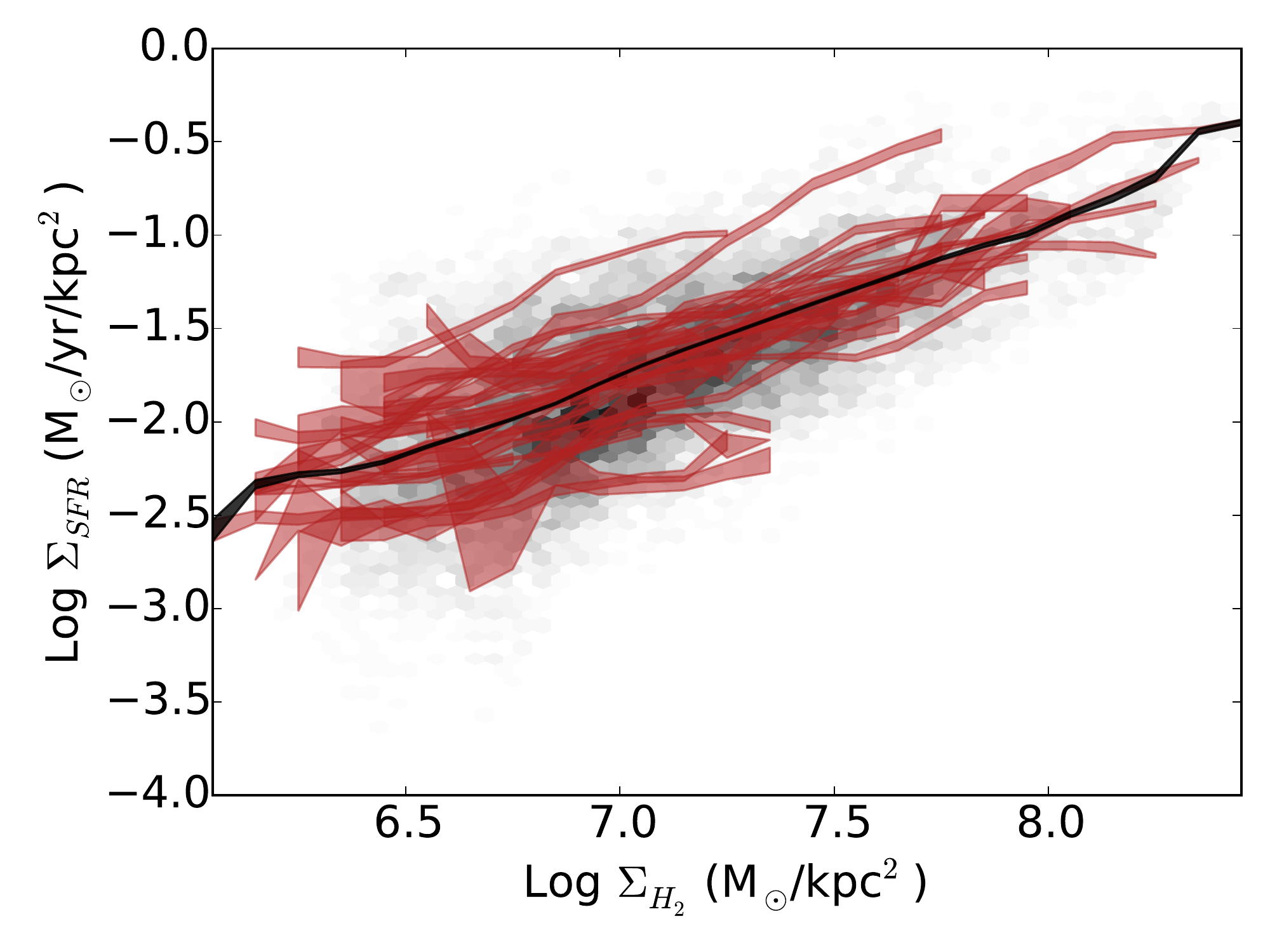}
	\includegraphics[width=5.5cm]{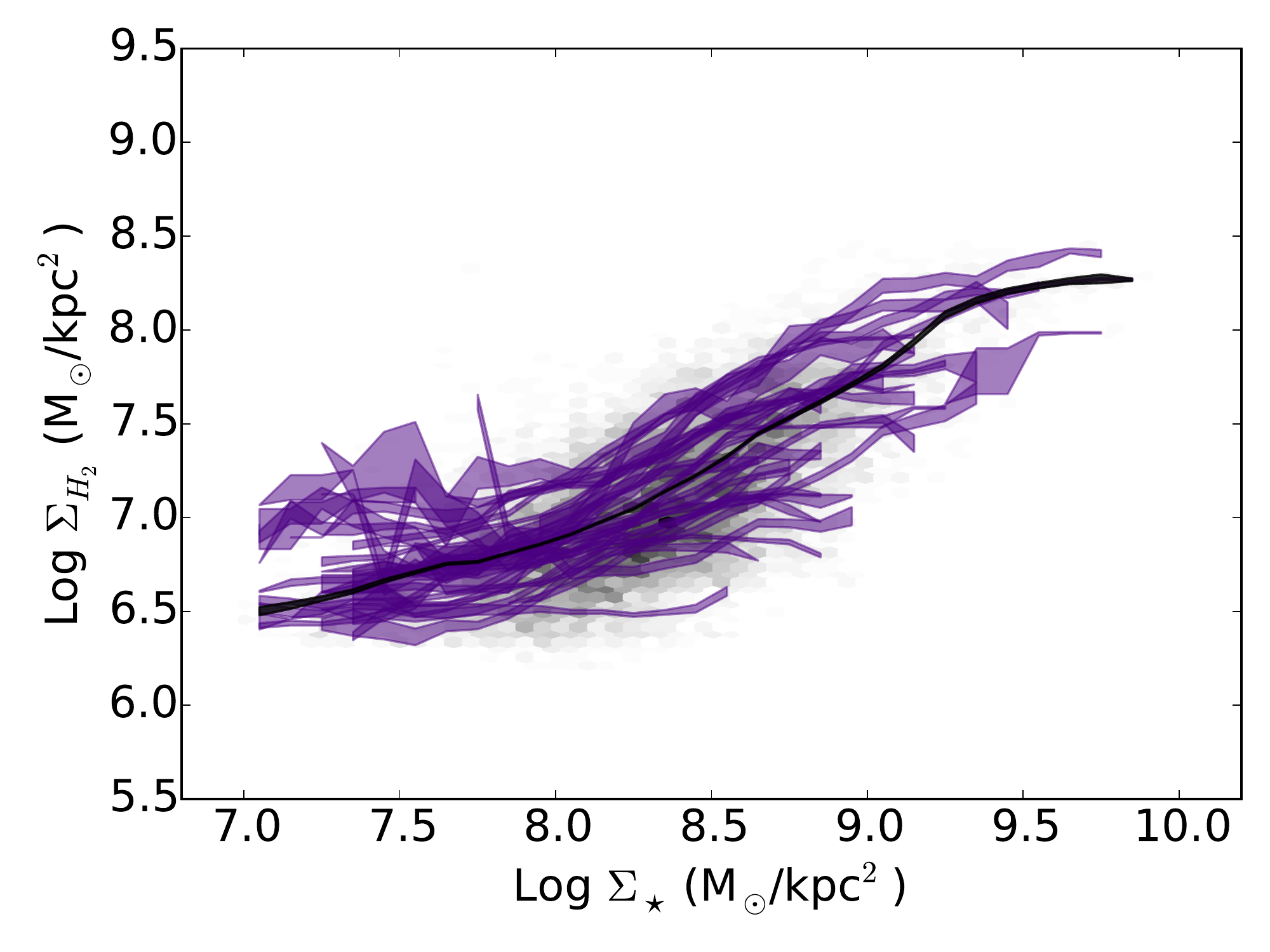}
        \caption{The galaxy-to-galaxy diversity of star formation scaling relations.  Left panel: Coloured curves show the median values of \sigsfr\ in bins of \sigstar\ for each of the 28 galaxies in our sample, and hence represent the rSFMS of each galaxy.  \sigstar\ bins have width of 0.3 dex and are offset by 0.1 dex to create a smoothed running median.  The width of each curve represents the vertical scatter of each scaling relation on a galaxy-by-galaxy basis and is calculated as $\sigma/\sqrt{N}$ for each x-axis position.  The greyscale background shows the number density for all $\sim$15,000 spaxels in the sample.  The black curve shows the running median for all $\sim$15,000 spaxels.  Middle panel:  As for the left panel, but for medians of \sigsfr\ as a function of \sigh2.  The coloured curves therefore represent the rSK relation for each galaxy and the black curve is the median of all $\sim$15,000 spaxels.  Right panel:  As for the left panel, but for medians of \sigh2\ as a function of \sigstar.  The coloured curves therefore represent the rMGMS for each galaxy and the black curve is the median of all $\sim$15,000 spaxels. }
    \label{curve_offsets}
\end{figure*}

Although all three resolved star formation scaling relations studied in this paper (the rSFMS, rSK relation and rMGMS) show significant galaxy-to-galaxy diversity, the visual impression from Figs \ref{rSFMS_fig}, \ref{rKS_fig} and \ref{rMGMS_fig} is that the relations show different amounts of variation (as also hinted by the different scatters in the ensemble relations in Fig. \ref{ensemble_fig}).  In this Section, we will quantify the galaxy-to-galaxy variation in the three star formation scaling relations, and compare it with the scatter of the full ensemble of $\sim$15,000 spaxels used in this work.  In the following Section, we will use the metrics developed here to investigate what galactic-scale parameters (if any) regulate the star formation relations and drive the galaxy-to-galaxy variations that are observed.

We begin with a further visual representation of the galaxy-to-galaxy variation in the star formation scaling relations.  Fig. \ref{curve_offsets} shows a compression of the information in Figs \ref{rSFMS_fig}, \ref{rKS_fig} and \ref{rMGMS_fig} into a single panel for each relation.  The grey scale background in each panel is once again the number density for the full sample of $\sim$15,000 spaxels (i.e. the same distributions as in Fig. \ref{ensemble_fig}) for reference.  The coloured curves in each panel represent a running median of the $y$-axis variable, in bins of 0.3 dex of the $x$-axis variable, offset by 0.1 dex along the $x$-axis in each step.    The black curve in each panel represents the running median of the full sample of $\sim$15,000 spaxels.  The width of each curve represents the vertical scatter of each scaling relation on a galaxy-by-galaxy basis and is calculated as $\sigma/\sqrt{N}$ for each x-axis position.  All three panels have the same dynamic range (4 dex) on the $y$-axis.

Fig. \ref{curve_offsets} reinforces the visual impression from  Figs \ref{rSFMS_fig}, \ref{rKS_fig} and \ref{rMGMS_fig} that the greatest galaxy-to-galaxy scatter is seen in the rSFMS, and least scatter in the rMGMS.  By comparing the coloured curves in each panel to the greyscale background, it can be appreciated that the scatter in any of the three ensemble relationships is encompassed by these galaxy-to-galaxy variations.

There are various methods that could be employed to quantify the galaxy-to-galaxy variation of the three star formation scaling relations.  Perhaps the simplest would be to sum (or average) spaxel values within a given galaxy in order to obtain the global (or average) sSFR, SFE and \fgas\ in each case.  This approach would capture a characteristic value of sSFR, SFE and \fgas\ for each galaxy that reflects a combination of the normalization and slope of the rSFMS, rSK relation and rMGMS.  However, summing spaxels in this way loses all of the benefit of a resolved study, and an equivalent investigation could be done with far superior statistics using surveys of global properties.

An alternative approach is to fit a first order polynomial to each relation in each galaxy and investigate the dependence of the normalization and slope on galaxy parameters of interest.  The main objection to this approach is that Figs \ref{rSFMS_fig}, \ref{rKS_fig} and \ref{rMGMS_fig} clearly demonstrate that the shapes of the star formation relations are very diverse.  Although the rSK relations in our data are largely well represented by a first order polynomial (which we will make use of later), this is not true for the rSFMS and rMGMS.  A higher order fit could of course be used (e.g. Hemmati et al. 2020), but it becomes non-trivial to compare the offset between galaxies with different shapes.

We opt for a hybrid of the above approaches, in which we quantify the offset between a given galaxy's scaling relation and the ensemble average of all spaxels in the sample, but without any parametric assumption on the shape of the relation within each galaxy.   Our approach is to compute the average offset between each of the coloured curves in a given panel of Fig. \ref{curve_offsets} and the black curve in the same panel.  Specifically, the offset is computed as the median difference between the black curve and a given coloured curve over all $x$-axis bins.  We can therefore compute, for each galaxy, a value of $\Delta$rSFMS, $\Delta$rSK and $\Delta$rMGMS, which is the median `offset' from a given scaling relation derived from all spaxels.  The uncertainty in $\Delta$rSFMS, $\Delta$rSK and $\Delta$rMGMS is given by the standard deviation of the differences between the black and coloured curves in each bin along the x-axis.  One drawback of this offset method is that it is sensitive to both changes in normalization and slope of a given relation, without discriminating between these effects explicitly.  

\begin{figure}
	\includegraphics[width=8.5cm]{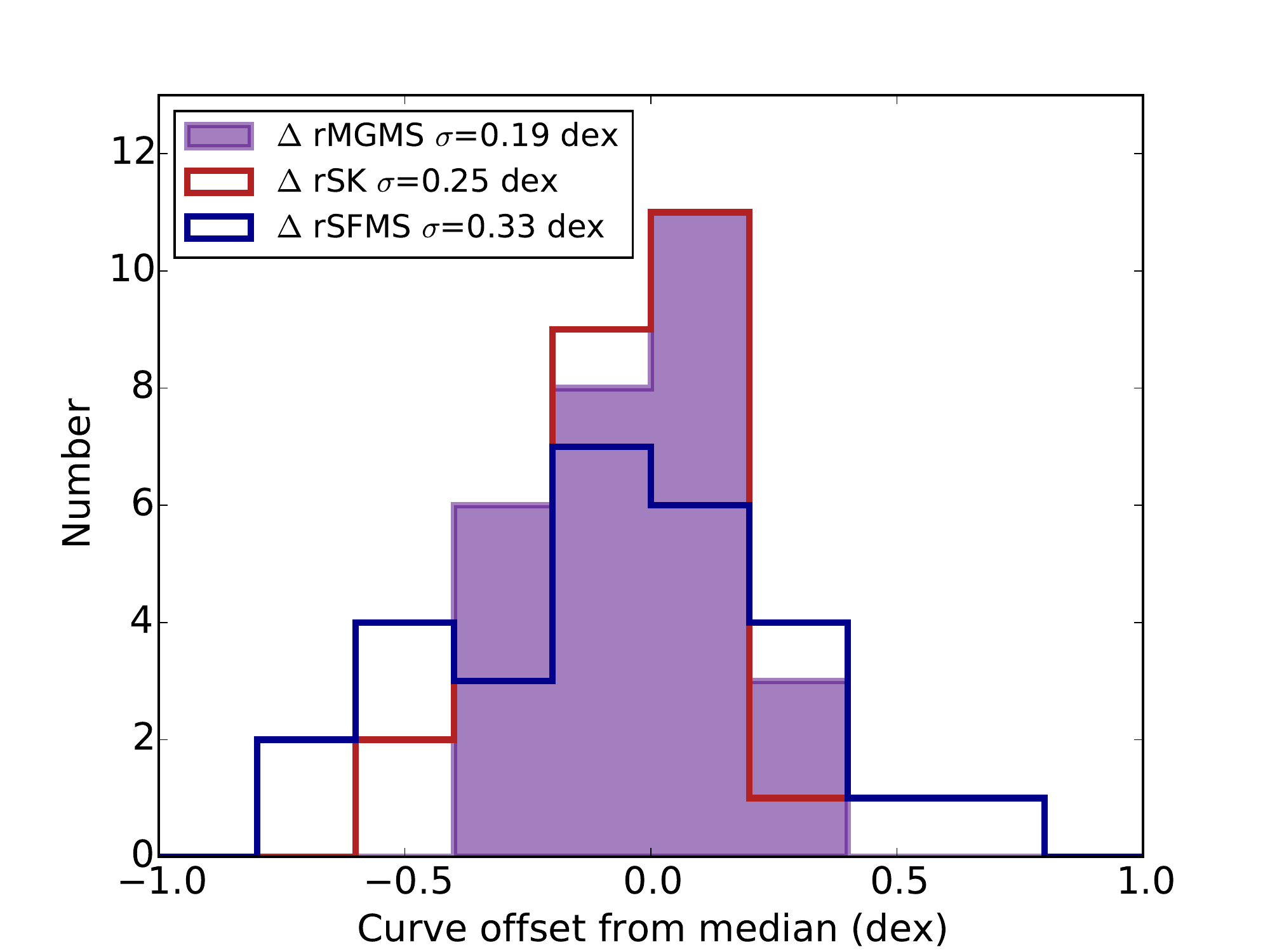}
        \caption{Histogram distribution of galaxy-by-galaxy offsets from the median scaling relations.  Offsets from the median rSFMS, rSK relation and rMGMS are shown in blue, red and purple respectively.  The galaxy-to-galaxy variation is greatest in the rSFMS and least in the rMGMS.}
    \label{curve_offset_hist}
\end{figure}

Fig. \ref{curve_offset_hist} shows the distribution of $\Delta$rSFMS, $\Delta$rSK and $\Delta$rMGMS amongst the 28 galaxies in our sample, with the RMS scatters reported in the legend.  Fig. \ref{curve_offset_hist} confirms our earlier visual impression from Figs \ref{rSFMS_fig}, \ref{rKS_fig} and \ref{rMGMS_fig} that the rSFMS shows the greatest galaxy-to-galaxy variation and that the rMGMS shows the least.  We note that our definition of $\Delta$rSFMS, $\Delta$rSK and $\Delta$rMGMS will not capture the full quantitative scatter of the ensemble relation, due to the variation in shapes and slopes from galaxy-to-galaxy.  For example, it is possible for a galaxy to have an offset of zero in one of the $\Delta$ quantities simply by virtue of a different slope.  It is therefore to be expected that the RMS values reported in  Fig. \ref{curve_offset_hist} differ from the scatters reported in Fig. \ref{ensemble_fig}.  Nonetheless, the distributions in  Fig. \ref{curve_offset_hist}, which show the smallest scatter for the rMGMS and the largest for the rSFMS, are qualitatively consistent with the ranking of scatters seen in the ensemble relations in Fig. \ref{ensemble_fig}.   We conclude that the rSFMS shows the greatest galaxy-to-galaxy variation (see also Lin et al. 2019) and the rMGMS is the most homogeneous of the three relations.

\section{Correlations between the star formation scaling relations}\label{co_sec}

\begin{figure}
	\includegraphics[trim=5cm 0cm 5cm 0cm,clip=true,width=9cm]{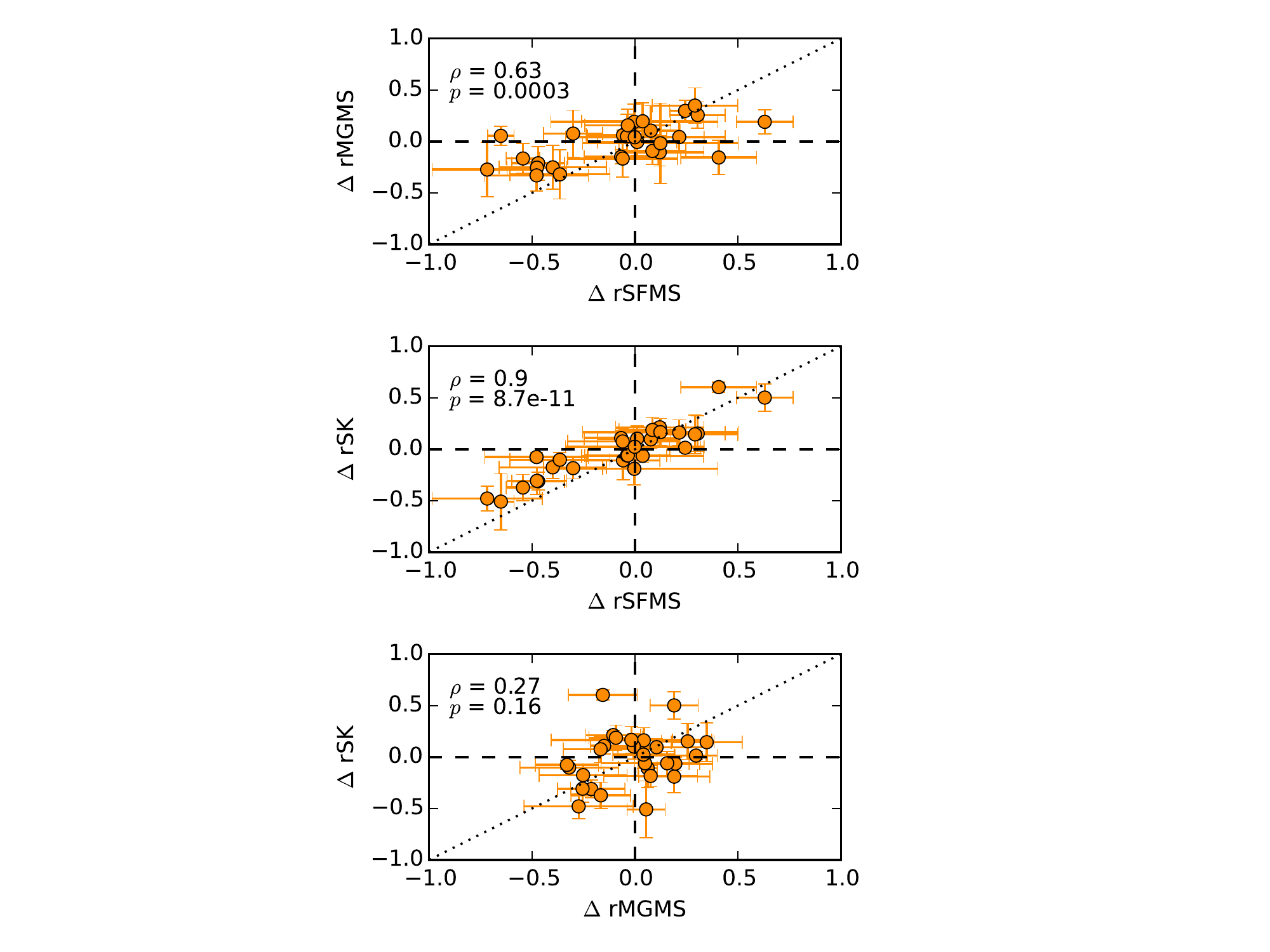}
        \caption{The galaxy-by-galaxy offset from each of the three star formation scaling relations is compared with offsets from the other relations.  The vertical and horizontal dashed lines show zero offset and the diagonal dotted line shows a one-to-one correlation.  The Pearson correlation coefficients ($\rho$) and $p$-values are reported in the top left corner of each panel. }
    \label{d_vs_d}
\end{figure}

Having defined a relative offset for each galaxy from the median of each of the three scaling relations, we can now investigate whether the offsets between different relations are correlated.  Such tests investigate whether the normalization of a given scaling relation tracks that of one of the other relations. 

In the lower panel of Fig. \ref{d_vs_d} we compare the offset of each galaxy from the rMGMS to its offset on the rSK relation.  No correlation is found between $\Delta$rMGMS and $\Delta$rSK, indicating that the response of the molecular gas distribution to the stellar potential, and the local recipe for the conversion of that gas into stars, are independent.  Indeed, Lin et al. (2019) found that including \sigstar\ as a variable in the parametrization of the rSK did not reduce its scatter.

Conversely, the middle panel of Fig. \ref{d_vs_d} shows that the offset of a given galaxy from the rSFMS is extremely well correlated with its offset from the rSK relation.  In addition to a very high Pearson correlation coefficient and miniscule $p$-value, the data points follow closely the one-to-one line.  Therefore, in contrast to offsets between the rSK relation and rMGMS (lower panel) \sigsfr\ seems to strongly link the change in rSFMS and the change in rSK relation on a galaxy-by-galaxy basis.  

In the top panel of Fig. \ref{d_vs_d} we show the last of the three combinations, the galaxy-by-galaxy offsets from the rSFMS and the rMGMS.   $\Delta$rSFMS and $\Delta$rMGMS have a significant correlation ($\sim$3.5$\sigma$), although this is not as strong as the correlation between $\Delta$rSFMS and $\Delta$rSK, whose significance is $\sim$7.5$\sigma$.  Galaxy-to-galaxy variation in the rSFMS therefore seems to be most closely linked to internal recipe for star formation, as set by the rSK relation.  This result is consistent with the work of Ellison et al. (2020a,b) in which we investigated offsets on a spaxel-by-spaxel basis and concluded that changes in \sigsfr\ around the rSFMS were more strong correlated with a spaxel's offset from the rSK than from the rMGMS, indicating that SFE, rather than gas fraction, is the more significant driver of scatter around the rSFMS.

There may be concern that correlations between variables, such as in Fig. \ref{d_vs_d}, may be artificially driven by `shared' parameters that appear on both axes.  However, as nicely illustrated by Dou et al. (2020), normalizing both axis variables by the same parameter (in their case, stellar mass) it is not trivial.  Moreover, because the variables plotted in Fig. \ref{d_vs_d} are offset ($\Delta$) values the parent variables do not even appear explicitly in either axis.  Indeed, there is no correlation at all between $\Delta$rMGMS and $\Delta$rSK (lower panel of Fig. \ref{d_vs_d}) despite the fact that \sigh2\ appears in both the rSK relation and the rMGMS.  The utility of the $\Delta$ metrics can be appreciated by considering a simple (albeit not realistic) scenario in which enhancements in \sigsfr\ are driven only by changes in the available gas reservoir - i.e. higher \sigh2\ at fixed \sigstar, but a normal rSK relation.  In such a scenario, we would measure a positive $\Delta$rSFMS and positive $\Delta$rMGMS, but $\Delta$rSK=0, i.e. no correlation between $\Delta$rSFMS and $\Delta$rSK, despite both scaling relations involving \sigsfr.  It is therefore possible to disentangle whether changes in \sigsfr\ at fixed \sigstar\ (i.e. non-zero $\Delta$rSFMS) are driven by changes in \sigsfr\ at fixed \sigh2\ (i.e. non-zero $\Delta$rSK, but normal rMGMS) or changes in \sigh2\ at fixed \sigstar\ (i.e. non-zero $\Delta$rMGMS, but normal rSK).  In practice, we find that both gas fractions and SFE play a role, since  $\Delta$rSFMS correlates with both $\Delta$rSK and $\Delta$rMGMS, but with a stronger effect by the former.  Although measurement errors could potentially cause correlations between offset variables, a Monte Carlo test of simulated data with the actual error distributions of our sample shows $\Delta$ values and correlation strengths that are much smaller than we observe (see the Appendix of Eales, Eales and de Vis 2020 for similar tests).  We conclude that the correlations shown in Fig. \ref{d_vs_d} are not due to shared variables or measurement errors.

Based on the larger scatter seen in the rSFMS (for their ensemble of galaxies) compared with the rMGMS or rSK relation, Lin et al. (2019) and Morselli et al. (2020) have recently argued that the rSFMS is a manifestation of combining the other two more `fundamental' relations.  Other recent works have demonstrated the tight inter-connectedness of stellar mass, molecular mass and SFR on the integrated level (Dou et al. 2020; Hunt et al. 2020).  Lin et al. (2019) proposed that the stellar potential sets the gas distribution, leading to the rMGMS, and the molecular gas distribution sets the rate of star formation, leading to the rSK relation.  Conversely, it is suggested that there is no fundamental physics linking \sigstar\ to \sigsfr, and that the rSFMS is the result of plotting two variables that are correlated for other reasons.  Fig. \ref{d_vs_d} offers complementary evidence for the conclusion that the rMGMS and rSK relations capture underlying physical processes whereas the rSFMS is the result of correlated variables.  Whereas the rMGMS and rSK relations for a given galaxy appear to be independent, the rSFMS of a given galaxy correlates significantly with both the rMGMS and the rSK relation.

Since many works (e.g. Leroy et al. 2013; Cano-Diaz et al. 2016, 2019; Barrera-Ballesteros et al. 2016; Hsieh et al. 2017; Vulcani et al. 2019; Sanchez 2020), including ours, have now shown that global galaxy scaling relations are rooted in local scale origins, we are led to the conclusion that the \textit{global} SFMS (as well as the rSFMS) is also a result of the co-variance at the local scale.  I.e. the global SFMS originates in the rSK and the rMGMS.  Indeed, Cicone et al. (2017) have shown that the correlation between total molecular gas mass and total stellar mass is as tightly correlated as the relation between SFR and molecular gas mass (see also Hunt et al. 2020).  The results of Cicone et al. (2017) therefore mirror our findings at the kpc-scale, that the \textit{global} MGMS is as tight as the \textit{global} SK relation, indicating an underlying physical connection between these variables.  Since the rSFMS is observed to exist even at relatively high redshift (e.g. Whitaker et al. 2012; Speagle et al. 2014; Popesso et al. 2019), the corollary of these conclusions is that the rMGMS and rSK relation were also in place at early times.

\section{What drives the galaxy-to-galaxy scaling relation differences?}\label{stuff_sec}

Having established that galaxies set their own internal star formation relations, we can now investigate whether there are identifiable global parameters that drive these differences.  Such drivers may differ for different scaling relations, and we must also be mindful that global parameters show inter-dependences which need to be disentangled (e.g. Teimoorinia et al. 2016; Bluck et al. 2020a).

\subsection{What drives the galaxy-to-galaxy variation of the resolved SFMS?}

\begin{figure}
	\includegraphics[trim={4cm 0 4cm 0},clip,width=10cm]{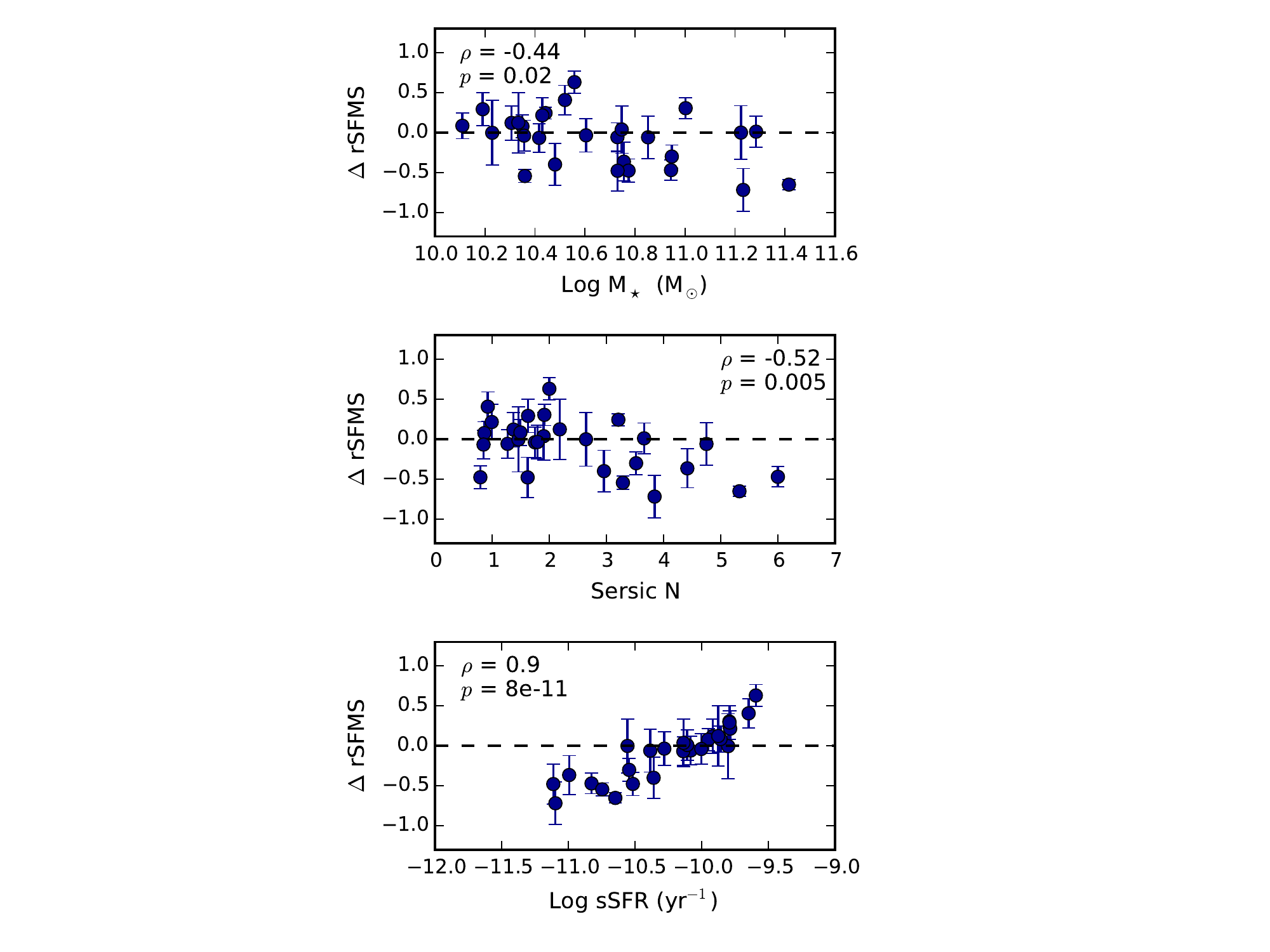}
        \caption{Galaxy-by-galaxy offset from the median rSFMS relation as a function of stellar mass (upper panel), Sersic index (middle panel) and sSFR (lower panel).  The Pearson correlation coefficient ($\rho$) and $p$-value are given in the corner of each panel. }
    \label{SFMS_stuff}
\end{figure}

\begin{figure}
	\includegraphics[width=8cm]{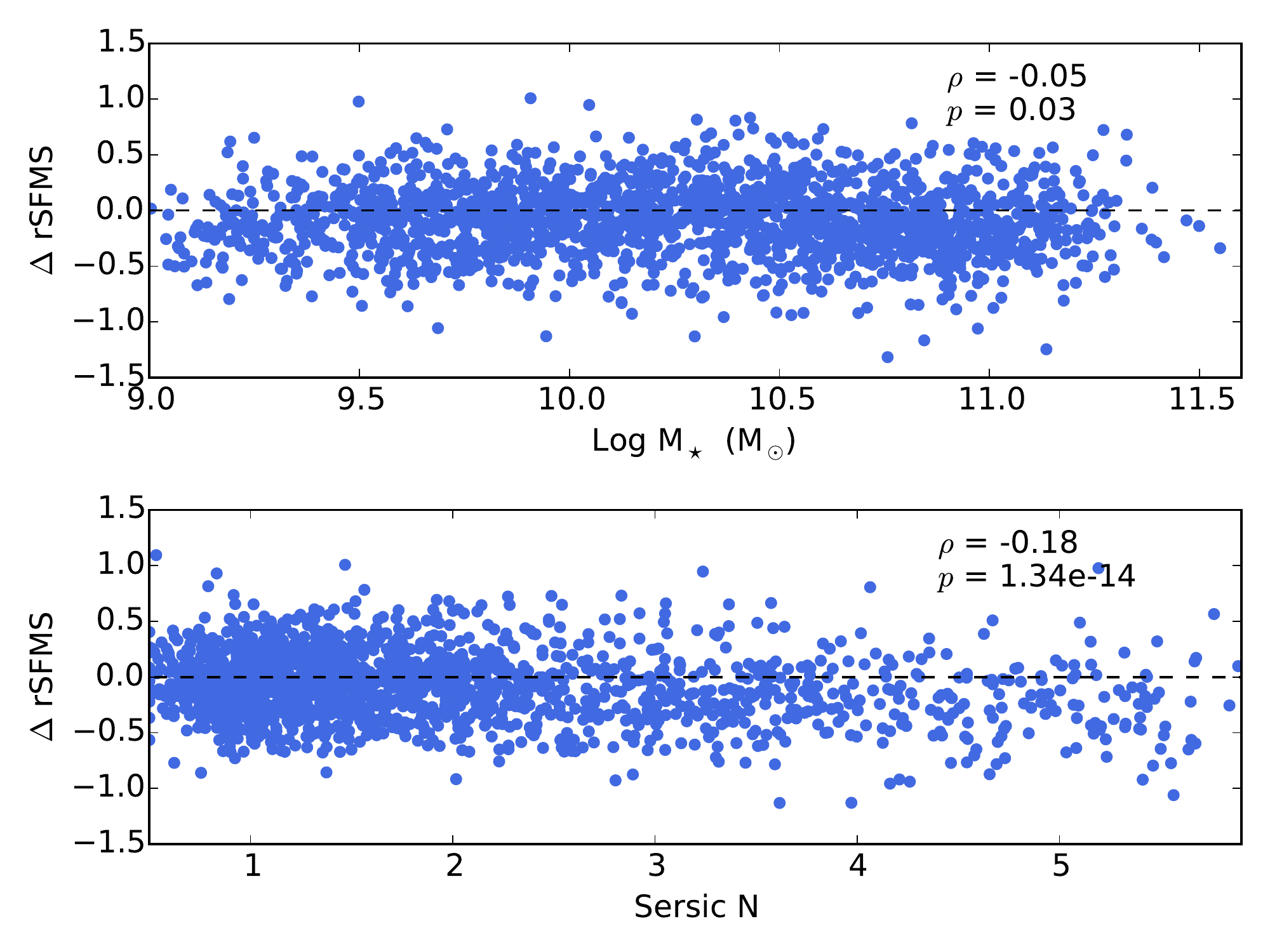}
        \caption{Galaxy-by-galaxy offset from the median rSFMS relation as a function of stellar mass (upper panel) and Sersic index (lower panel) for $\sim$ 2000 galaxies selected from MaNGA DR15.  The Pearson correlation coefficient ($\rho$) and $p$-value are given in the top right of each panel. With this larger sample, the correlation between rSFMS and Sersic becomes highly significant.}
    \label{dSFMS_all}
\end{figure}

In Fig. \ref{SFMS_stuff} we plot the offset of each galaxy's rSFMS from the median relation (i.e. black curve in the top panel of Fig. \ref{curve_offsets}) versus a variety of galactic properties: total stellar mass, Sersic index (taken from the NSA catalog and the best morphology metric available for all of the galaxies in our sample) and global sSFR.  The Pearson correlation coefficient  ($\rho$) and $p$-value are reported in the corner of each panel.

The strongest correlation in Fig. \ref{SFMS_stuff} is between a given galaxy's offset from the median rSFMS and its total sSFR.  However, this is a trivial result, since the sSFR of a galaxy is essentially the same measure as rSFMS offset, since it measures the summed \sigsfr\ relative to the summed \sigstar.  The only reason that the correlation between $\Delta$rSFMS and sSFR is not perfect is because the sSFR is taken from the PIPE3D VAC, rather than computed directly from the \sigsfr\ measurements computed by us.  Moreover, the PIPE3D VAC computes its sSFR by summing the H$\alpha$ fluxes from all spaxels, whereas we have used a set of very specific cuts to define star forming spaxels, as appropriate for the purposes of this paper.  Despite the trivial nature of the $\Delta$rSFMS -- sSFR relation, we have included it here for consistency with the tests made in the next two subsections for the rSK relation and the rMGMS.

The strongest (non-trivial) relation in Fig. \ref{SFMS_stuff} is the anti-correlation between $\Delta$rSFMS and the Sersic index (N$_s$) of a given galaxy, in the sense that more galaxies with a classic elliptical profile (N$_s$=4) tend to have a rSFMS below the average.    Disk-like morphologies (N$_s \sim 1$) lie close to, or above the average rSFMS of the sample.  The anti-correlation in our data is present at $\sim 3\sigma$ significance.  Our result reflects trends previously seen in the global SFMS, where earlier type morphologies are offset to low SFRs for their mass (e.g. Wuyts et al. 2011; Catalan-Torrecilla et al. 2017; Sanchez 2020).  Our results also confirm previous studies of the rSFMS that have found a morphology dependence (e.g. Gonzalez-Delgado et al. 2016; Maragkoudakis et al. 2017; Medling et al. 2018; Pan et al. 2018).  In the most recent and extensive investigation of this dependence, Cano-Diaz et al. (2019) have shown a steady decrease in the characteristic \sigsfr\ at fixed \sigstar\ for earlier morphologies, also in agreement with our results.

There is a weaker, but still significant ($p=0.02$) anti-correlation between $\Delta$rSFMS and total stellar mass.  A similar (mild) dependence of the rSFMS on stellar mass was also reported by Bolatto et al. (2017) for the EDGE-CALIFA sample.  A qualitatively similar trend between stellar mass and rSFMS offset has also been found in simulations (Trayford \& Schaye 2019), albeit with a larger magnitude and stronger significance than found in our \aq\ data.  However, we propose that any mass dependence of the rSFMS (at least in the observational data) is likely driven by the fact that morphology is a function of total stellar mass, with more massive galaxies tending to be more bulge dominated (e.g. Kelvin et al. 2014; Thanjavur et al. 2016; Bottrell et al. 2017; Bluck et al. 2019).  Indeed, there is a strong correlation ($\rho$=0.8) between total stellar mass and Sersic index in our sample.  

Given the relatively small size of our sample, we test whether the correlations are strongly driven by one of the data points. To achieve this, we follow a jack-knife approach in which each galaxy is removed in turn from the sample and the $p$-value is recomputed.  We then count the fraction of these jack-knife iterations for which a correlation of at least 2$\sigma$ significance is detected ($p<0.05$).  Unsurprisingly, the strong correlations between $\Delta$rSFMS and both sSFR and Sersic index are very robust, and 100 per cent of the jack-knife iterations yields a significant (by our 2$\sigma$ definition) correlation.  The weaker correlation between $\Delta$rSFMS and stellar mass is still recovered with at least 2$\sigma$ significance for 93 per cent of the jack-knife iterations.  We conclude that the correlations in Fig. \ref{SFMS_stuff} are not driven by a single outlier.

Since investigation of the rSFMS does not require matched ALMA data, we can further test the correlation (and robustness thereof) between $\Delta$rSFMS and Sersic index and stellar mass by using the MaNGA DR15 sample as a whole.  We repeat our basic galaxy and spaxel selection as described in Section \ref{data_sec} (the galaxy must have $b/a\ge0.35$, and have at least 20 star-forming spaxels to define its rSFMS) to identify 1944 galaxies containing a total of 1.4 million spaxels.  The ensemble rSFMS is defined using this expanded set of 1.4 million star forming spaxels and offsets are computed on a galaxy-by-galaxy basis as described above.  The results are shown in Fig. \ref{dSFMS_all}.  The weakly significant anti-correlation between total stellar mass (top panel) is still present with $p=0.03$, but is visually not convincing.  However, the anti-correlation between $\Delta$rSFMS and Sersic index is now highly significant ($p \sim 10^{-14}$).  With this larger sample, it is also evident that the morphology dependence of rSFMS is driven by galaxies with N$_s>$3. These high Sersic index galaxies represent only 15 percent of the $\sim$2000 galaxies selected from DR15, but 35 per cent of our ALMaQUEST sample, hence the latter is able to reveal the significant dependence of rSFMS on N$_s$, even with its modest sample size.

\begin{figure}
	\includegraphics[width=9cm]{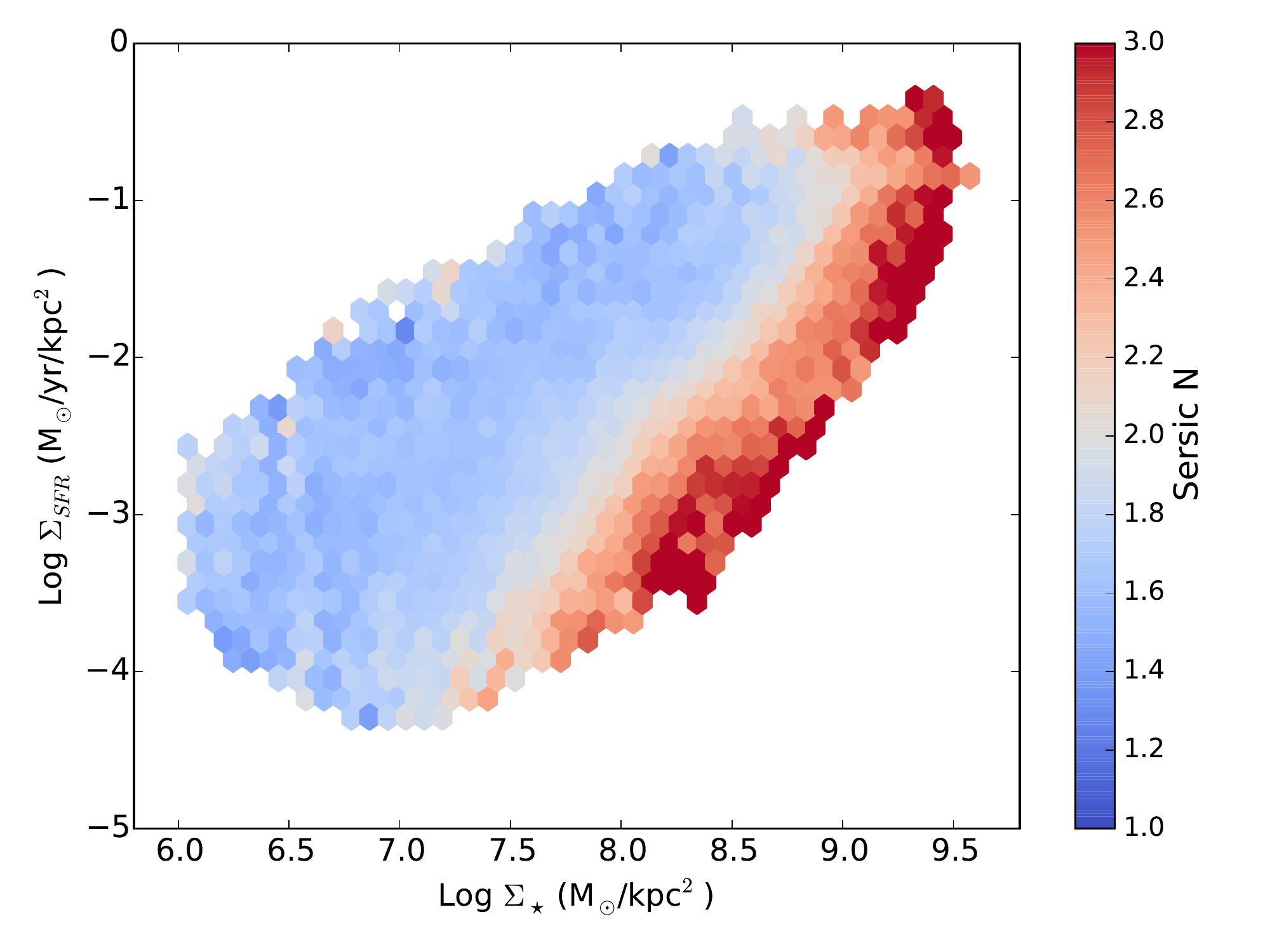}
        \caption{The rSFMS for 1.4 million spaxels from $\sim$ 2000 MaNGA DR15 galaxies.  Each binned element is colour coded by the median Sersic index of its host galaxy.  At fixed \sigstar\ the lowest \sigsfr\ values are in high Sersic index galaxies.}
    \label{SFMS_N}
\end{figure}

A final demonstration of the dependence of the rSFMS on N$_s$ for 1.4 million star-forming spaxels in MaNGA DR15 is presented in Fig. \ref{SFMS_N}, in which we plot the rSFMS, colour coding each binned element by the median N$_s$ of the host galaxy from which each spaxel is drawn.  It is clearly seen that at fixed \sigstar\ the lowest \sigsfr\ values are located in galaxies with higher Sersic indices.  However, there is considerable scatter in $\Delta$rSFMS even at fixed Sersic index, consistent with the results of Enia et al. (2020), who find variation in the rSFMS even within their sample of local disk galaxies.

\subsection{What drives the galaxy-to-galaxy variation of the resolved SK relation?}

\begin{figure}
	\includegraphics[trim={4cm 0 4cm 0},clip,width=10cm]{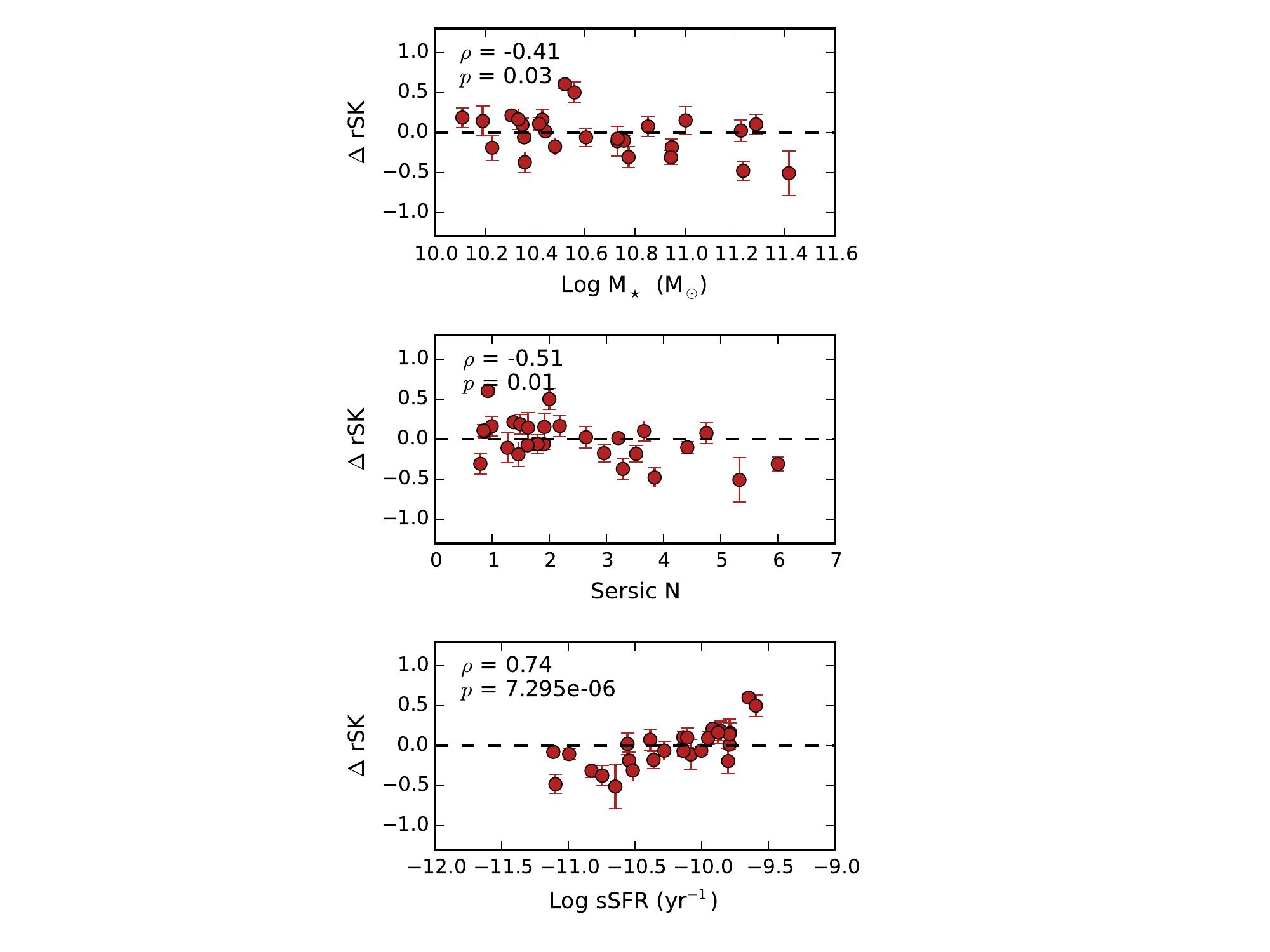}
        \caption{Galaxy-by-galaxy offset from the median rSK relation as a function of stellar mass (upper panel), Sersic index (middle panel) and sSFR (lower panel).  The Pearson correlation coefficient  ($\rho$) and $p$-value are given in the top left of each panel. }
    \label{KS_stuff}
\end{figure}

\begin{figure}
	\includegraphics[trim={4cm 0 4cm 0},clip,width=10cm]{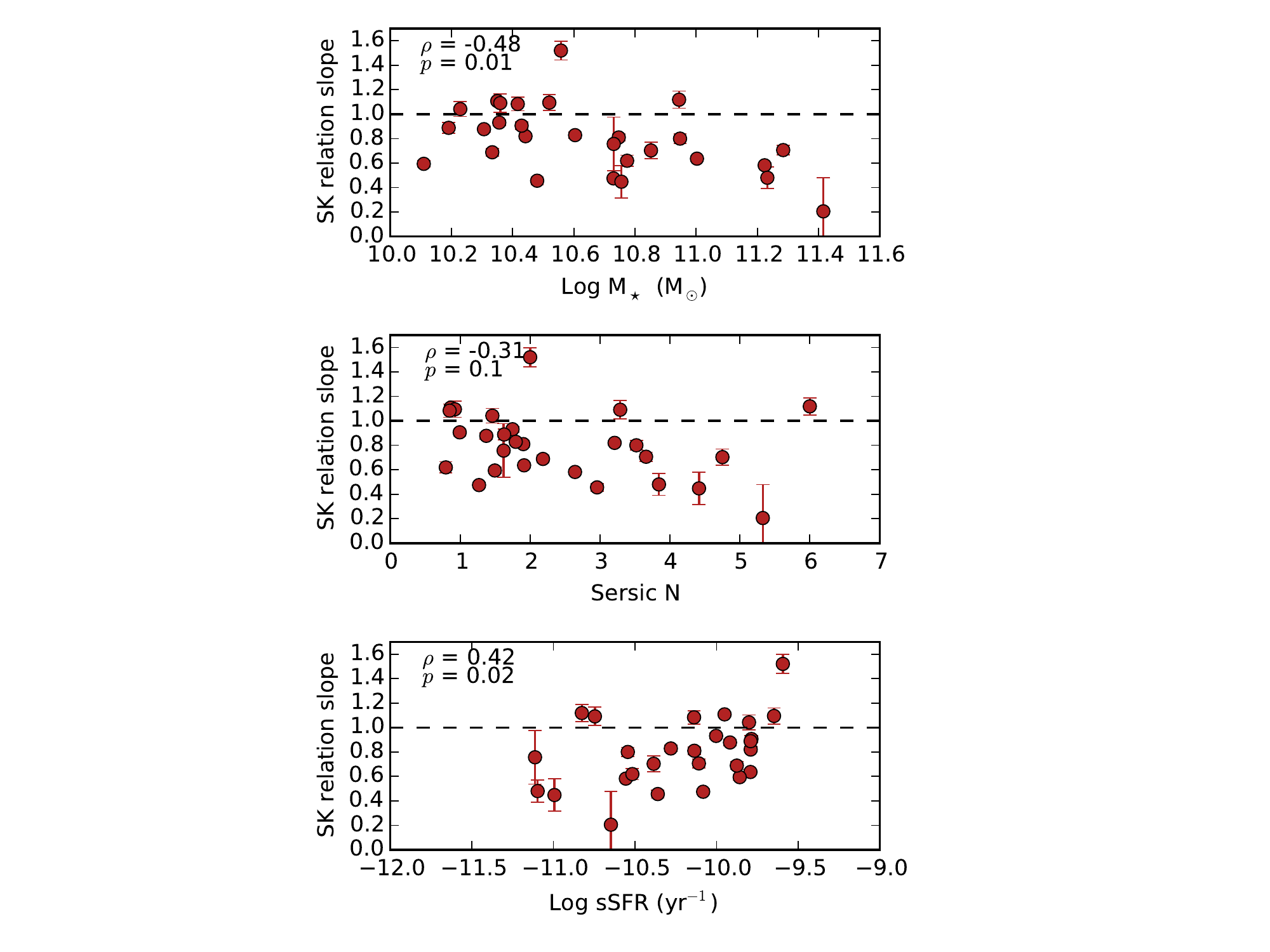}
        \caption{Galaxy-by-galaxy slope of the rSK relation as a function of stellar mass (upper panel) and Sersic index (middle panel) and sSFR (lower panel).  The Pearson correlation coefficient ($\rho$) and $p$-value are given in the top left of each panel. }
    \label{KS_slopes}
\end{figure}

\begin{figure}
	\includegraphics[trim={4cm 0 4cm 0},clip,width=10cm]{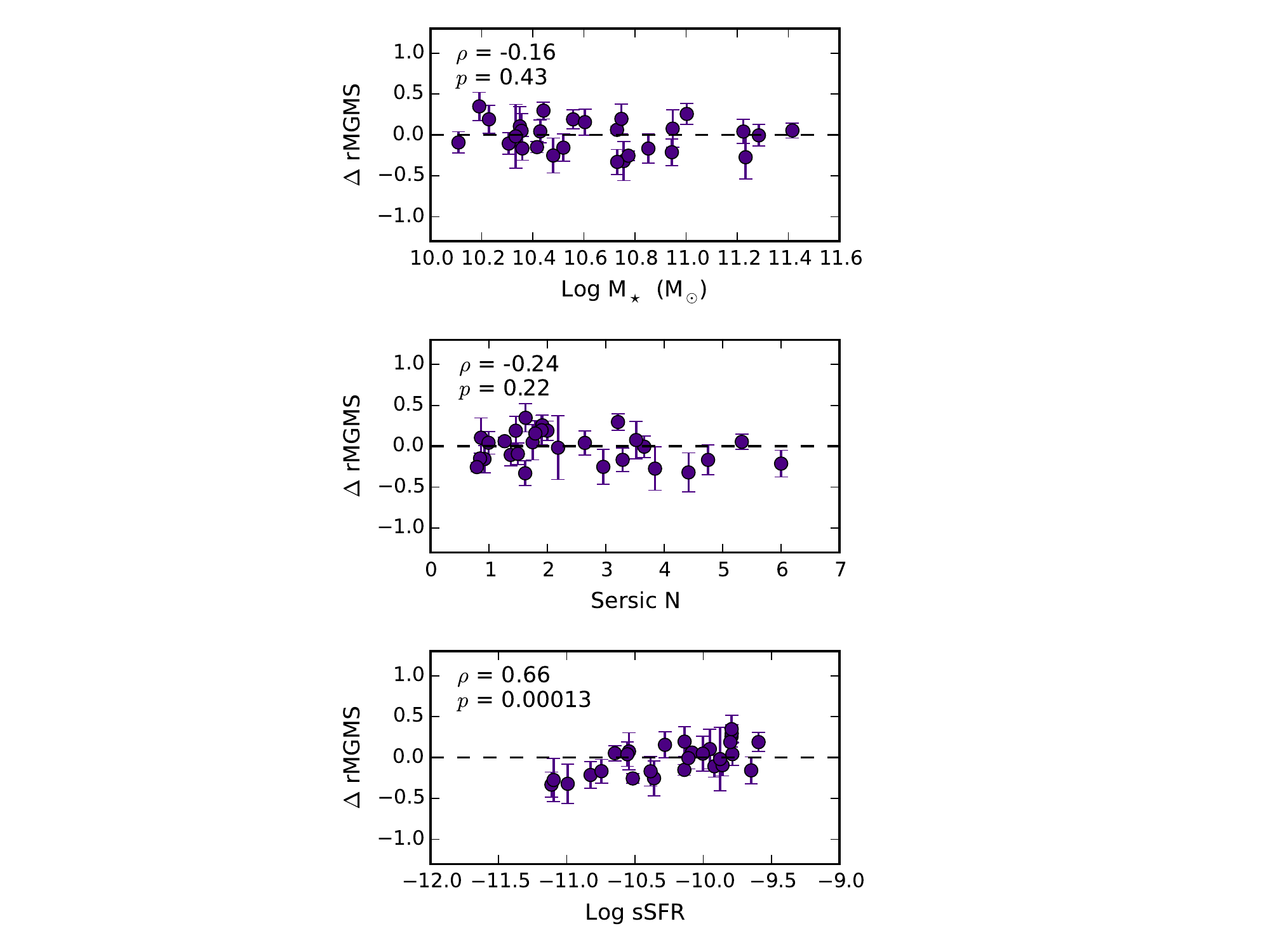}
        \caption{Galaxy-by-galaxy offset from the median rMGMS as a function of stellar mass (upper panel),  Sersic index (middle panel) and sSFR (lower panel).  The Pearson correlation coefficient ($\rho$) and $p$-value are given in the top left of each panel. }
    \label{MGMS_stuff}
\end{figure}

In Fig. \ref{KS_stuff} we plot the dependence of the galaxy-by-galaxy offset from the rSK relation on stellar mass (upper panel), Sersic index (middle panel) and sSFR (lower panel).   A strong and highly significant correlation ($\rho \sim 0.7$ and $p=7\times10^{-6}$) is found between the offset of a galaxy's rSK relation and its global sSFR (lower panel of Fig. \ref{KS_stuff}).   Although SFR contributes to both axes in this relation (either as \sigsfr\ or total SFR), we include the trend of $\Delta$rSK with sSFR here in order to be able to compare to published analyses of these same quantities, either on resolved or global galaxy data.  Moreover, as we have seen throughout this paper, variables that indirectly contribute to both axes do not always lead to correlations (e.g. lower panel of Fig. \ref{d_vs_d}).  The implication of the correlation between sSFR and $\Delta$rSK is that when a galaxy shows an elevated SFR for its stellar mass, the cause is an elevated SFE.   Leroy et al. (2013) found a similar correlation between the normalization of the rSK law (as parametrized by the integrated depletion time in their study) and sSFR for a sample of 30 local galaxies with kpc-scale data.  Ellison et al. (2020b) likewise found that kpc-scale spaxels in the \aq\ sample showed elevated SFRs primarily as a result of enhanced SFE.  

A similar conclusion has also been reached for global measurements of molecular gas in local galaxies, which find a strong link between depletion time (i.e. the inverse of SFE as defined here) and sSFR (Saintonge et al. 2011b; Huang \& Kauffmann 2014; Bolatto et al. 2017).  Indeed, Dou et al. (2020) have claimed that the sSFR-SFE relation is the most fundamental, from which other scaling laws derive.  The correlation between $\Delta$rSK and sSFR is indeed the strongest of the relations reported here.  Taken together, these results show that the global dependence of SFE on sSFR in SDSS galaxies, like many global galaxy correlations, derives from local-scale relationships (as shown analytically, and for the general case, by Sanchez-Almeida \& Sanchez-Menguiano 2019).  High redshift galaxies exhibit the same dependence on main sequence offset and depletion time (e.g. Genzel et al. 2015; Silverman et al. 2018; Tacconi et al. 2018), with recent studies showing considerable complexity in their inter-correlations (Liu et al. 2019).  Nonetheless, main sequence galaxies appear to define a broadly consistent global SK relation from $z=0$ to at least $z\sim4$, but with starbursts offset above it (e.g. Daddi et al. 2010; Bethermin et al. 2015; Schinnerer et al. 2016). It therefore seems plausible that the high redshift rSK relation, and its dependence on sSFR (or main sequence offset) will also be driven by kpc-scale dependences.

Also in agreement with resolved (Leroy et al. 2013; Bolatto et al. 2017; Utomo et al. 2018; Zabel et al. 2020) and global (Saintonge et al. 2011b, 2012; Bothwell et al. 2014) studies of depletion time (or SFE, or SFE per free-fall time), we additionally find a weak anti-correlation ($\sim 2\sigma$) between $\Delta$rSK and total stellar mass.  Indeed, it is interesting to note that the correlation coefficients for our resolved relations are very similar to those for the global relations (see Fig. 3 of Saintonge et al. 2011b).  However, we additionally find that there is a significant anti-correlation ($p=0.01$) between $\Delta$rSK and Sersic index (middle panel of Fig. \ref{KS_stuff}) that is stronger than the relation with total stellar mass.  In a complementary study, Colombo et al. (2018) also found a dependence of SFE on morphology, as parametrized by Hubble type (see also Sanchez 2020).    As noted in the previous subsection, given that there is a strong correlation between total stellar mass and Sersic index, we speculate that the correlation between $M_{\star}$ and $\Delta$rSK may therefore be rooted in an underlying dependence on morphology.  Indeed, a dependence of the star formation efficiency on varying morphology indicators (such as Hubble type, or concentration parameter) has been found for both resolved (e.g. Leroy et al. 2013) and global (e.g. Saintonge et al. 2012) studies, and SFEs are lower in ellipticals than spirals (Davis et al. 2014).  In a complementary machine learning approach, Dey et al. (2019) have also found that star formation on kpc-scales is suppressed in larger, earlier type galaxies.  These observational results are replicated by simulations, in which larger bulge fractions lead to more stable disks and hence a reduction in SFE (Gensior et al. 2020).

Visually, it can be seen from Fig. \ref{rKS_fig} that (in log-log space) the rSK relation in the galaxies in our sample can be well represented by a simple first order polynominal.   In addition to the parametrization of $\Delta$rSK, we can therefore also investigate the variation in the rSK relation slopes from galaxy-to-galaxy.   A slope of unity indicates that a given galaxy forms stars with the same efficiency at all values of \sigh2.  Such an investigation is less straightforward for the rSFMS and the rMGMS, whose shapes show significant variation from galaxy-to-galaxy (e.g. Figs. \ref{rSFMS_fig} and \ref{rMGMS_fig}). 

We determine the slope of the galaxy-by-galaxy rSK relation using a least squares fit to the individual spaxel data, with errors determined directly from the covariance matrix of the fit (see Leroy et al. 2013 for a discussion of how weighting and fit methods can affect fit values)\footnote{An ODR fit was also performed, but produced erratic results for a handful of galaxies with small numbers of spaxels, or limited dynamic range.  The OLS fit was also a better visual representation of the data for the galaxy-by-galaxy fits and is hence adopted here.}.  The slopes of the rSK relation for each galaxy in our sample are mostly in the range 0.4 to 1.2, with a median value of 0.86 that is (unsurprisingly) consistent with the slope of the ensemble of all the spaxels in the sample (Fig. \ref{ensemble_fig}).  The rSK relation slopes in our sample are generally somewhat flatter than those in the HERACLES sample (e.g. Fig. 5 of Leroy et al. 2013), but as shown in that paper (and here) the nature of these relations depend on the galaxy properties of the sample. There are two notable outliers in the distribution of rSK relation slopes in our sample:  8082-12704 whose rSK slope is 0.2 and 8081-3704 whose rSK slope is 1.5.  The former of these outliers has relatively few star forming spaxels with which to constrain its rSK relation, and they are all at quite large galactocentric radii (e.g. Fig \ref{rKS_fig}).  We conclude that the rSK relation fit for 8082-12704 is likely to be unreliable.  Conversely, the rSK relation for 8081-3704 seems to be robust, and it is obvious from Fig \ref{rKS_fig} that its slope is steep.  This galaxy is one of the central starbursts studied by Ellison et al. (2020a), and was identified in that work to have one of the largest enhancements in central SFE of the sample, consistent with our finding of a steep rSK relation.  

In Fig.  \ref{KS_slopes} we plot the slope of the rSK relation versus the same global galaxy parameters that we have investigated for a correlation with the normalization of the resolved star formation relations.   The horizontal dashed line shows a slope of unity, for reference.  From Fig. \ref{KS_slopes} we find correlations of at least 2$\sigma$ significance between the slope of the rSK relation and sSFR (correlation) and with total stellar mass (anti-correlation).  These trends with total stellar mass and sSFR were also seen in Fig. \ref{KS_stuff} to correlate with the offset from median rSK relation, although the trends with rSK slope are weaker than with $\Delta$rSK.  In contrast, whereas $\Delta$rSK shows a significant correlation with Sersic index, no significant correlation is found between N$_s$ and the slope of the rSK relation.  The weaker trends seen between sSFR, total stellar mass and N$_s$ with rSK slope, indicates that the normalization of the rSK relation is affected by these galaxy parameters, in addition to the slope.

Given the varying molecular gas detection limits from galaxy-to-galaxy (e.g. Fig. \ref{rKS_fig}) we repeat the above analysis but requiring a minimum log \sigh2 = 7.0 M$_{\odot}$ kpc$^{-2}$ for a spaxel to be included in the analysis.  Since all of the galaxies in our sample have \sigh2\ detection thresholds below log \sigh2 = 7.0 M$_{\odot}$ kpc$^{-2}$, this experiment enforces a uniform effective surface density threshold.  The results of Fig. \ref{KS_stuff} are qualitatively unchanged with the imposition of this surface density threshold.  The significance of the correlation between $\Delta$rSK and both stellar mass and Sersic index are identical to those shown in Fig. \ref{KS_stuff}.  The correlation between $\Delta$rSK is actually strengthened with the molecular gas surface density cut imposed: $\rho$=0.81, $p$=3.2$\times 10^{-7}$.  However the correlation between galaxy physical properties and SK relation slope (Fig. \ref{KS_slopes}) is affected by the imposition of a detection threshold: the correlation with sSFR is no longer statistically significant ($\rho$=0.26, $p$=0.2) and the correlation with stellar mass is weaker, but still significant above the 2$\sigma$ threshold considered in this work ($\rho$=$-0.42$, $p$=0.03).

As in the previous subsection, we test whether the correlations in Figs \ref{KS_stuff} and \ref{KS_slopes} are driven by a single galaxy by performing a jack-knife test.  As before, each galaxy is removed in turn and the $p$-value is re-computed.  A correlation of at least 2$\sigma$ significance between $\Delta$rSK and both Sersic index and sSFR is recovered for all jack-knife iterations, as well as in 93 per cent of iterations with stellar mass.  Likewise, the two significant correlations with SK slope (sSFR and stellar mass) are recovered in 96 per cent of the iterations.  We conclude that a single outlier does not drive the correlations in Figs. \ref{KS_stuff} and \ref{KS_slopes}.

\subsection{What drives the galaxy-to-galaxy variation of the resolved MGMS?}

Fig. \ref{MGMS_stuff} examines the correlations between a galaxy's offset from the average rMGMS and global parameters.  In contrast to the rSK relation and rSFMS, no significant correlation is found between $\Delta$rMGMS and either Sersic index or total stellar mass (despite the fact that stellar mass contributes to the definition of $\Delta$rMGMS, highlighting again that correlations found in previous sections are not a trivial result of shared variables).  The rMGMS of a galaxy correlates only (within the parameters tested here) on its sSFR ($\sim 3.5\sigma$ significance).   The lack of sensitivity of the rMGMS to total stellar mass and Sersic index likely contributes to its relatively small scatter overall.

Global studies of molecular gas content have found a strong correlation of \fgas\ with sSFR (Saintonge et al. 2011a, 2017; Dou et al. 2020), which is consistent with our finding of a highly significant correlation of this latter variable with $\Delta$rMGMS.  Saintonge et al. (2017) have additionally found an anti-correlation between \fgas\ and both total stellar mass and morphology (parametrized by global stellar mass surface density in their work), which we do not find in our resolved data.  However, the trends between \fgas\ and stellar mass and morphology were found by Saintonge et al. (2017) to be mass dependent and require stacking to properly account for non-detections, whereas we have \textit{required} CO detections for at least 20 spaxels for galaxies to be included in our sample.  The fairest comparison of our work is therefore probably with the individual detections of galaxies in the original COLDGASS survey (Saintonge et al. 2011a), that span a very similar mass range to our sample.  When only the COLDGASS detections are considered, no significant correlation is found between \fgas\ and either total stellar mass or morphology, consistent with our findings in Fig \ref{MGMS_stuff}.

As in the previous subsection, we test the impact on our results imposing a uniform detection threshold for all spaxels of log \sigh2 = 7.0 M$_{\odot}$ kpc$^{-2}$.  The statistically significant correlation found in the lower panel of Fig \ref{MGMS_stuff} between $\Delta$rMGMS and sSFR is still present, but with a slightly weaker significance ($\rho=0.6$, $p$=0.0013).  Likewise, our jack-knife test reveals that the statistically significant correlation between $\Delta$rMGMS and sSFR remains in 100 per cent of iterations, indicating that neither the detection threshold nor the a single outlier is responsible for the correlation.

\medskip

In summary of this Section, we have found the following (non-trivial) correlations to exist between the three star formation scaling laws and galactic parameters with at least 2$\sigma$ significance:

\begin{itemize}
\item   Anti-correlation between $\Delta$rSFMS and total stellar mass ($p=0.02$) and Sersic index ($p=0.005$).
\item   Anti-correlation between $\Delta$rSK and total stellar mass ($p=0.03$) and Sersic index ($p=0.01$); Correlation between $\Delta$rSK and sSFR ($p=7\times10^{-6}$).
\item   Correlation between $\Delta$rMGMS and sSFR ($p=0.0001$).
\end{itemize}

\section{Discussion}\label{discuss_sec}

We have presented a large sample of $\sim$15,000 spaxels in 28 galaxies observed with MaNGA and ALMA to homogeneously derive \sigsfr, \sigstar\ and \sigh2\ in order to investigate the rSFMS, rMGMS and rSK relation on a galaxy-by-galaxy basis.  All three relations show considerable galaxy-to-galaxy variation in both normalization and shape.  We also find that galaxies show an order of magnitude internal variation in their sSFR, SFE and \fgas, as explored for the full \aq\ sample in Lin et al. (2020).

\subsection{Internal variations and self-similar disks.}

Although future papers in this series will analyse radial properties, as well as the behaviour of gas and star formation in the bulge and disk regions, we comment briefly on this topic in the context of the results presented here.  Radial gradients (and variations therein) in sSFR, SFE and \fgas\ have all been extensively documented in previous works (Leroy et al. 2008, 2013; Schruba et al. 2011; Cano-Diaz et al. 2016; Casasola et al. 2017; Ellison et al. 2018; Sanchez et al. 2018; Sanchez 2020), so the existence of internal variations of these properties is not surprising.  However, our results contribute to this discussion by showing that galaxies are clearly not self-similar, even at fixed total stellar mass, or at fixed galactocentric radius.

The variation of the rSFMS from galaxy-to-galaxy provides a good example of this lack of self-similarity and its relevance in current literature debates.  On the one hand, it is not suprising that the shape (both a sub-unity slope and a flattening of \sigsfr\ at high \sigstar) of the rSFMS indicates that the sSFR is lower in the centre for many galaxies in our sample.  After all, the fundamental distinction in colour between galactic bulges and disks tells us that star formation dominates in the latter.    Abramson et al. (2014) have indeed argued that separating the global SFMS into a bulge and disk component leads to a more constant sSFR for the disk, and hence the slope of the global SFMS is driven by the relative contribution of the bulge.  Mendez-Abreu, Sanchez \& de Lorenzo-Caceres  (2019) have also found that the integrated SFRs in the disk components of early type galaxies are qualitatively consistent with the global SFMS.

However, the diversity in the rSFMS amongst galaxies shown in Fig. \ref{rSFMS_fig} is clearly not in agreement with a constant disk rSFMS, which would manifest as a universal relation (with constant normalization) between \sigstar\ and \sigsfr\ on disk scales (i.e. above a few kpc).  Instead, we find that there is a range of characteristic sSFRs on disk scales even for galaxies with approximately constant total stellar mass.  The slopes of the rSFMS in this radius regime are also generally sub-unity.  Our results are therefore more consistent with the recent work of Cook et al. (2020) who find that the reduced sSFR of the global SFMS at high M$_{\star}$ persists in the disks of galaxies.  Catalan-Torrecilla et al. (2017) likewise find disks that can be offset below the SFMS.  We also find a variety of behaviours for the central (or bulge) regions in our sample - often the \sigsfr\ turns over at high \sigstar, but in some cases, the relation continues as a straight line.  This is again in agreement with Catalan-Torrecilla et al. (2017) who find a wide range of bulge sSFR.  In short, we do not find a universal disk and bulge rSFMS that simply vary in their relative proportions from galaxy to galaxy.   The scaling relations found in this work appear to be neither universal, nor self-similar at a given radial distance.  

\subsection{Interpreting star formation relations on different scales.}

As emphasized in the Introduction, scaling relations in science are highly valued for the insight that they provide into the fundamental mechanisms that drive physical processes.  The scaling relations presented in this paper were first established for integrated galaxy properties, such as total stellar mass, total SFR and total gas content metrics (e.g. Kennicutt 1998a,b; Noeske et al. 2007a).  The existence of such tight, global scaling relations has been bestowed with significant fundamental import in understanding the regulation of star formation (e.g. Noeske et al. 2007b; Renzini \& Peng 2015), and has been used as a metric for the successful reproduction of galactic properties in cosmological simulations (e.g. Sparre et al. 2015; Matthee \& Schaye 2019)\footnote{Whilst the rSK relation is imposed manually in many simulations as a star formation recipe, the rSFMS is not an explicit input and its existence is therefore a result of a combination of structure formation and resulting star formation.}.

Spatially resolved studies have now firmly established that the star formation scaling relations reflect kpc-scale correlations (e.g. Wong \& Blitz 2002; Bigiel et al. 2008; Schruba et al. 2011; Cano-Diaz et al. 2016; Hsieh et al. 2017), and simulations have likewise tested whether resolved relations emerge from contemporary hydrodynamical codes (e.g. Orr et al. 2018; Trayford \& Schaye 2019; Hani et al. 2020).  Attention has therefore naturally shifted to parametrizing the shape, slope and normalization of resolved relations, in order to understand, for example, whether depletion times are constant.  As a result, physical significance has been attached to the fits derived from kpc-scale surveys of gas and star formation.  However, the significant galaxy-to-galaxy variations in all three of the star formation scaling relations undermines attempts to make universal statements based on ensembles of data.  This caveat was pointed out several years ago already by Leroy et al. (2013) for the rSK relation, who cautioned that to fit a single power law to the ensemble data of many galaxies oversimplifies their actual complexity.  The details of the fit to any data ensemble will depend on the  exact galaxies that are included, and therefore carries little physical meaning.  For example, Leroy et al. (2013) have pointed out that the rSK relation exhibits an `artificial' uniformity when the sample is dominated by star-forming disks.  Resolution can also affect the derived slope (Calzetti et al. 2012), as well as conversions between different molecular gas tracers (e.g. different excitation CO lines).

The work presented here demonstrates that the same caveat applies to all three of the resolved star formation scaling relations.  Perhaps the most dramatic demonstration of the perils of fitting a single relation to a diverse population is the rSFMS.  Fig. \ref{rSFMS_fig} shows that the individual galaxy rSFMS often turnover towards high \sigstar.  As discussed above, this is a natural consequence of lower sSFR in the centres of many (but not all) galaxies.  However, the ensemble rSFMS for all of the $\sim$ 15,000 spaxels in our dataset, shows no such turnover (Fig. \ref{ensemble_fig}) - the detailed shapes of individual rSFMS have been smeared out when combined into a single sample.  

Pan et al. (2018) have previously proposed that a turnover in the rSFMS at high \sigstar\ is the result of an increasing contribution from non-star forming regions in the bulge.  They find that a linear rSFMS is recovered when only strictly star forming spaxels are selected from the MaNGA dataset.  Our results differ from those of Pan et al. (2018), since we find that individual galaxies may exhibit  a turnover in their rSFMS even though we have applied a strict set of criteria to identify star formation dominated spaxels.  We suggest that the likely origin of our differing conclusions is that the turnover of the rSFMS that exists in individual galaxies is erased when galaxies are compiled together.  In practice, both effects are likely at play, i.e. the centres of galaxies contain fewer star forming regions, \textit{and} there is a real suppression in the centres of galaxies.  A complementary result is presented by Cano-Diaz et al. (2019) who show that galaxies with different morphologies have different rSFMS, not just because they have fewer regions of star formation, but also because there is a different characteristic normalization.

Comparing the diverse shapes and normalizations from galaxy-to-galaxy in Fig. \ref{rSFMS_fig} and the ensemble of all the spaxels in Fig. \ref{ensemble_fig} also highlights the challenge of interpreting the shape of the global SFMS from large datasets.  For example, Renzini \& Peng (2015) have emphasized the tight, linear ridge line of the global SFMS obtained from the SDSS DR7, and discussed the previous observations of flattening at high M$_{\star}$ (e.g. Whitaker et al. 2012) in terms of bias in sample selection.  However, our results indicate that diversity of the galaxy-to-galaxy rSFMS can be lost when large ensembles of galaxies are combined, an effect which fundamentally undermines the extraction of physical meaning from the shape of global relations.

Despite the new insights that can be gleaned from moving from studying the star formation relations on a global scale to kpc-scales, there is further progress to be made by continuing our journey down the resolution scale.  The rSK relation is known to breakdown on scales below a few hundred pc (e.g. Onodera et al. 2010; Schruba et al. 2010), with a spatial mis-match between the locations of gas and stars (Kreckel et al. 2018; Kruijssen et al. 2019; Schinnerer et al. 2019).  The resolved kpc-scale relations studied in this paper therefore only emerge when the resolution is sufficient to blur these locations into overlap.  As a result, the scatter of the rSK relation exhibits a clear dependence on resolution (Kreckel et al. 2018).  Whilst we have therefore cautioned that we must be mindful of over-interpreting either global scaling relations, or ensembles of data from many galaxies, the kpc-scale relations for individual galaxies may also mask the fundamental physics of the star formation process.


\subsection{What drives the variation in scaling relations?}

Studies of global gas properties in large galaxy samples indicate that variations in SFR are driven by changes in both gas fraction and SFE (Tacconi et al. 2013, 2018; Piotrowska et al. 2020).  Similar investigations have also been conducted using kpc-scale IFU datasets (e.g. Dey et al. 2019; Bluck et al. 2020a,b; Morselli et al. 2020).  In Ellison et al. (2020b) we used the \aq\ dataset to quantify whether the rSFMS was primarily regulated by changes in gas fraction or SFE.  Using both a traditional correlation analysis, as well as an artificial neural network approach, Ellison et al. (2020b) concluded that whilst both gas fractions and SFE were linked to changes in the rSFMS, it was the latter that exhibited the stronger correlation.  The results in Fig. \ref{d_vs_d} support the conclusion of Ellison et al. (2020b).  Although there is a significant correlation between a galaxy's rMGMS (which reflects its characteristic gas fraction) and the rSFMS (top panel), the relationship is much stronger with the rSK relation (an indicator of SFE; middle panel).  Nonetheless, it is important to note that this generalized conclusion for our dataset does not necessarily apply to every individual galaxy, and it is almost certainly the case that some galaxies have sSFRs that are responding to gas supply.  As shown in Fig. \ref{profiles_fig}, the cause of reduced central sSFRs can vary, being due to either low gas fraction or low SFE, further highlighting the individuality of every galaxy.

In this paper, we have focused largely on the differences in the star formation relations from galaxy-to-galaxy.  Other works have investigated differences \textit{within} a given galaxy.  Leroy et al. (2013) were amongst the first to demonstrate significant galaxy-to-galaxy differences in the rSK relation, but they also demonstrated that a given galaxy tends to show shorter depletion times in its centre.  With expanded samples it has become clear that the situation is more complicated.  For example, using the EDGE-CALIFA sample Colombo et al. (2018) showed that radial depletion time profiles depend strongly on morphology. Also with EDGE-CALIFA, Utomo et al. (2017) have demonstrated a large diversity in radial depletion times; like Leroy et al. (2013), they identified a subset of galaxies that have shorter depletion times in their centres.  Galaxies with strong bars, or recent interactions often fall into this category of short central depletion times (although not all barred galaxies have shorter central depletion times; Chown et al. 2019).  However, Utomo et al. (2017) also showed that some galaxies can have central depletion times that are longer than the galaxy average.   Once again, the galaxy population exhibits considerable diversity, not just between, but within galaxies, even on kpc scales.  

\subsection{Extra pieces of the star formation puzzle.}

Despite our efforts to bring together the ingredients for the three star formation relations, the picture afforded by the \aq\ dataset remains incomplete.  We are lacking both atomic gas, as well as the dense gas (most commonly traced by HCN) measurements, both of which have been recently shown to play an important role in understanding the balance of gas and star formation.  The role of HI in regulating star formation has sometimes been down-played since early observations, with molecular gas shown to drive the rSK relation on both global and resolved scales (Wong \& Blitz 2002; Bigiel et al. 2008; de los Reyes \& Kennicutt 2019).  However, Morselli et al. (2020) have recently shown (for a sample of five local grand design spirals with kpc-scale resolution) that \sigsfr\ can be expressed solely as a function of \sigstar\ and $\Sigma_{\rm HI}$, with the former variable setting the rSFMS and the latter its normalization (see also Saintonge et al. 2016 for a global perspective on the dependence of the rSFMS and HI content).  Casasola et al. (2020) have also shown that (globally) dust mass correlates better with HI than molecular gas and Bacchini et al. (2019) find a strong correlation between HI and SFR when volume based (rather than surface density) quantities are used.  A more extensive analysis of the role of $\Sigma_{\rm HI}$ in regulating star formation in the nearby universe likely awaits the arrival of the ngVLA or SKA.

Yet perhaps the most interesting relationship that we have not been able to study in the current work is the correlation between the dense gas surface density ($\Sigma_{\rm HCN}$ and \sigsfr, or `dense gas rSK relation').  This relation, due to its small scatter and consistency from the scales of molecular clouds to galaxies, has been proposed as a demonstration that star formation is a simple function of the dense gas content (Wu et al. 2005; Lada et al. 2010).  However, recent work from the EMPIRE survey has shown that even the dense gas rSK relation is not immune from significant galaxy-to-galaxy, as well as internal, variations (Gallagher et al. 2018; Jimenez-Donaire et al. 2019).  For example, these works show that although dense gas is generally more abundant in the centres of galaxies, the efficiency with which this gas is turned into stars is low at smaller galactocentric radii (see also earlier work by Usero et al. 2015; Bigiel et al. 2016).  Mapping HCN at kpc-scale resolution for MaNGA galaxies is now within the reach of the ALMA observatory, and in Lin et al. (in prep) we present detections for two main sequence and 3 green valley galaxies in our sample.

\section{Summary and conclusions}\label{summary_sec}

Large statistical surveys of galaxies have demonstrated that there are tight correlations between total stellar mass, total molecular gas mass, and total SFR, which have become known as the Schmidt-Kennicutt (SK) relation (M$_{H2}$ - SFR), the star forming main sequence (SFMS; M$_{\star}$ - SFR) and the molecular gas main sequence (MGMS; M$_{\star}$ - M$_{H2}$).  These same star formation scaling relations have been shown to exist on local (kpc or below) scales within galaxies, leading to the definition of a `resolved' SK relation (rSK; \sigh2\ - \sigsfr), a `resolved' SFMS (rSFMS; \sigstar\ - \sigsfr) and a `resolved MGMS' (rMGMS; \sigstar\ - \sigh2).  The resolved star formation relations have been demonstrated to exist, with a tight scatter, for large ensembles of galaxies, broadly indicating that galaxies are built and regulated according to these physical parameters.

Using a sample of $\sim$15,000 star forming spaxels selected from 28 galaxies in the \aq\ survey, we have investigated the galaxy-to-galaxy variation of the three resolved star formation relations (rSK, rSFMS, rMGMS) in order to assess their universality and correlations with global galaxy parameters.  Our main conclusions are as follows:

\begin{enumerate}
  
\item  Based on our full sample of $\sim$15,000 star forming spaxels, we confirm the strong kpc-scale correlations found in many previous works between \sigh2\  and \sigsfr\ (i.e. the rSK relation),  \sigstar\ and \sigsfr\ (i.e. the rSFMS) and between \sigstar\ and \sigh2 (i.e. the rMGMS).  The scatter of the rSK relation and the rMGMS is equally tight ($\sim0.2$ dex for an ODR fit, or $\sim0.3$ dex for an OLS fit), and their correlation coefficients equally strong ($\rho \sim$ 0.7).  However, the rSFMS shows both a higher scatter (0.3 dex for the ODR and 0.4 dex for the OLS fits) and weaker correlation coefficient ($\rho$= 0.57), see Fig. \ref{ensemble_fig}.

  \medskip
  
\item  Despite the strong, and tight, correlations observed in the ensemble relations, there is up to an order of magnitude variation in all three of the star formation relations between different galaxies which dominates the scatter of the ensemble relations (Fig. \ref{curve_offsets}).

  \medskip

\item The median offset of a given galaxy from each of the three star formation scaling relations is computed relative to the median of the full sample of $\sim$15,000 spaxels in order to yield a $\Delta$rSK, $\Delta$rSFMS and $\Delta$rMGMS for each galaxy.  A comparison of these galaxy-by-galaxy offsets reveals that the rMGMS exhibits the least galaxy-to-galaxy variation and the rSFMS shows the most (Fig. \ref{curve_offset_hist}).

  \medskip

\item  There is no correlation between a galaxy's $\Delta$rSK and its $\Delta$rMGMS, indicating that the distribution of molecular gas in the stellar potential and its conversion into stars is regulated independently within each galaxy.  However, there is a strong correlation between $\Delta$rSFMS and both $\Delta$rSK and $\Delta$rMGMS (Fig. \ref{d_vs_d}).  We suggest that this is a result of the rSFMS arising from the combination of the rSK relation and rMGMS, rather than being an independent correlation in its own right.  The existence of the rSFMS as a product of the the rMGMS and rSK relation is consistent with its larger scatter and weaker correlation coefficient.

  \medskip

\item  We investigate how $\Delta$rSK, $\Delta$rSFMS and $\Delta$rMGMS depend on global galaxy properties and find the following (non-trivial) correlations with at least 2$\sigma$ significance:  the rSFMS is lower in galaxies with higher M$_{\star}$ and larger Sersic index (Fig. \ref{SFMS_stuff}, \ref{dSFMS_all}, \ref{SFMS_N}); the rSK is lower in galaxies with higher M$_{\star}$, larger Sersic index and lower sSFR (Fig. \ref{KS_stuff}); the rMGMS is lower in galaxies with lower sSFR (Fig. \ref{MGMS_stuff}).  We also find that the slope of the rSK relation correlates with sSFR and anti-correlates with stellar mass (Fig. \ref{KS_slopes}).

  \medskip

  
\end{enumerate}

A full understanding of what drives these galaxy-to-galaxy variations remains elusive.  One promising recent result in this regard is the finding that using volumetric versions of the scaling relations significantly reduces their scatter, since it accounts for the variations in scale height and effects such as disk flaring (Bacchini et al. 2019).  Datasets that include multiple gas phase tracers (such as HI, H$_2$ and HCN), as well as computing volumetric scaling relations may shed further light on the magnitude of galaxy-to-galaxy variation of the star formation scaling laws and the factors that drive them.

\section*{Acknowledgements}

The authors thank the referee for a constructive and helpful report.  SLE gratefully acknowledges support from an NSERC Discovery Grant.  LL thanks support from the following grants: Academia Sinica under the Career Development Award CDA-107-M03 and the Ministry of Science \& Technology of Taiwan under the grant MOST 107-2119-M-001-024 - and 108-2628-M-001 -001 -MY3.  We gratefully acknowledge grant MOST 107-2119-M-001-024 for funding travel to ASIAA (SLE and MDT).  SFS  thanks the following projects for support: CONACYT  FC-2016-01-1916 and CB-285080 and PAPIIT IN100519.  AFLB and RM acknowledges ERC Advanced Grant 695671 ”QUENCH” and support by the Science and Technology Facilities Council (STFC).  We thank Viviana Casasola, Ryan Chown, Maria Jesus Jimenez-Donaire and Laura Morselli for useful comments on an earlier draft of this paper. This paper makes use of the following ALMA data: ADS/JAO.ALMA\#2015.1.01225.S, ADS/JAO.ALMA\#2017.1.01093.S, ADS/JAO.ALMA\#2018.1.00558.S, ADS/JAO.ALMA\#2018.1.00541.S.   ALMA is a partnership of ESO (representing its member states), NSF (USA) and NINS (Japan), together with NRC (Canada), MOST and ASIAA (Taiwan), and KASI (Republic of Korea), in cooperation with the Republic of Chile. The Joint ALMA Observatory is operated by ESO, AUI/NRAO and NAOJ.  The National Radio Astronomy Observatory is a facility of the National Science Foundation operated under cooperative agreement by Associated Universities, Inc.

The SDSS is managed by the Astrophysical Research Consortium for the Participating Institutions. The Participating Institutions are the American Museum of Natural History, Astrophysical Institute Potsdam, University of Basel, University of Cambridge, Case Western Reserve University, University of Chicago, Drexel University, Fermilab, the Institute for Advanced Study, the Japan Participation Group, Johns Hopkins University, the Joint Institute for Nuclear Astrophysics, the Kavli Institute for Particle Astrophysics and Cosmology, the Korean Scientist Group, the Chinese Academy of Sciences (LAMOST), Los Alamos National Laboratory, the Max-Planck-Institute for Astronomy (MPIA), the Max-Planck-Institute for Astrophysics (MPA), New Mexico State University, Ohio State University, University of Pittsburgh, University of Portsmouth, Princeton University, the United States Naval Observatory, and the University of Washington.

\section*{Data Availability}

The MaNGA data cubes used in this work are publicly available at https://www.sdss.org/dr15/.  The ALMA data used in this work are publicly available after the standard one year proprietary period via the ALMA archive: http://almascience.nrao.edu/aq/.

\appendix

\section{Conversion factor concerns}\label{alpha_sec}

The choice of $\alpha_{CO}$ is the bane of molecular gas studies.  In this work, we have adopted a constant value of $\alpha_{CO}$ = 4.3 M$_{\odot}$ pc$^{-2}$ (K km s$^{-1}$)$^{-1}$.  However, it is well known that this value decreases when the star formation rate is high, as first found for extreme starburst galaxies (Downes \& Solomon 1998; Bryant \& Scoville 1999), and later as a smooth function of offset above the global SFMS (Accurso et al. 2017).  Although the radial profiles of $\alpha_{CO}$ are generally flat within a given galaxy, galaxies frequently show decreased values within the inner 1 kpc (Sandstrom et al. 2013).  Finally, due to a combination of few metal atoms per hydrogen, as well as the altered ionization structure due to reduced shielding at lower metallicities and dust-to-gas ratios, there is also an inverse correlation between  $\alpha_{CO}$ and metallicity (e.g. Narayanan et al. 2012).  We refer the reader to Bolatto et al. (2013) for an extensive review on variations in the CO conversion factor.

We investigate the impact of a metallicity dependent $\alpha_{CO}$ on our results by adopting the parametrization from Sun et al. (2020):

\begin{equation}\label{alpha_eqn}
  \alpha_{CO} = 4.35 Z^{-1.6} M_{\odot} pc^{-2} (K~km~s^{-1})^{-1}
\end{equation}

where $Z$ is the spaxel gas-phase metallicity computed using the Pettini \& Pagel (2004) calibration, normalized to a solar value of 12+log(O/H)=8.69.  A S/N$>$5 in each of the calibration's emission lines is required. Of the $\sim$15,000 spaxels in our sample, 10,562 pass this criterion and hence have metallicity dependent  $\alpha_{CO}$ computed.  The median change in \sigh2\ when adopting equation 1 is $0.07$ dex, i.e. on average \sigh2\ is slightly larger for the sample overall using the metallicity dependent conversion factor (since the median metallicity is slightly sub-solar at 12+log(O/H)=8.65.  The majority of metallicities are between 8.55 $<$ 12+log(O/H) $<$ 8.75, resulting in conversion factors that are typically between 3.5 $<$ $\alpha_{CO}$ $<$ 7.3 M$_{\odot}$ pc$^{-2}$ (K km s$^{-1}$)$^{-1}$.).

\begin{figure*}
	\includegraphics[width=5.5cm]{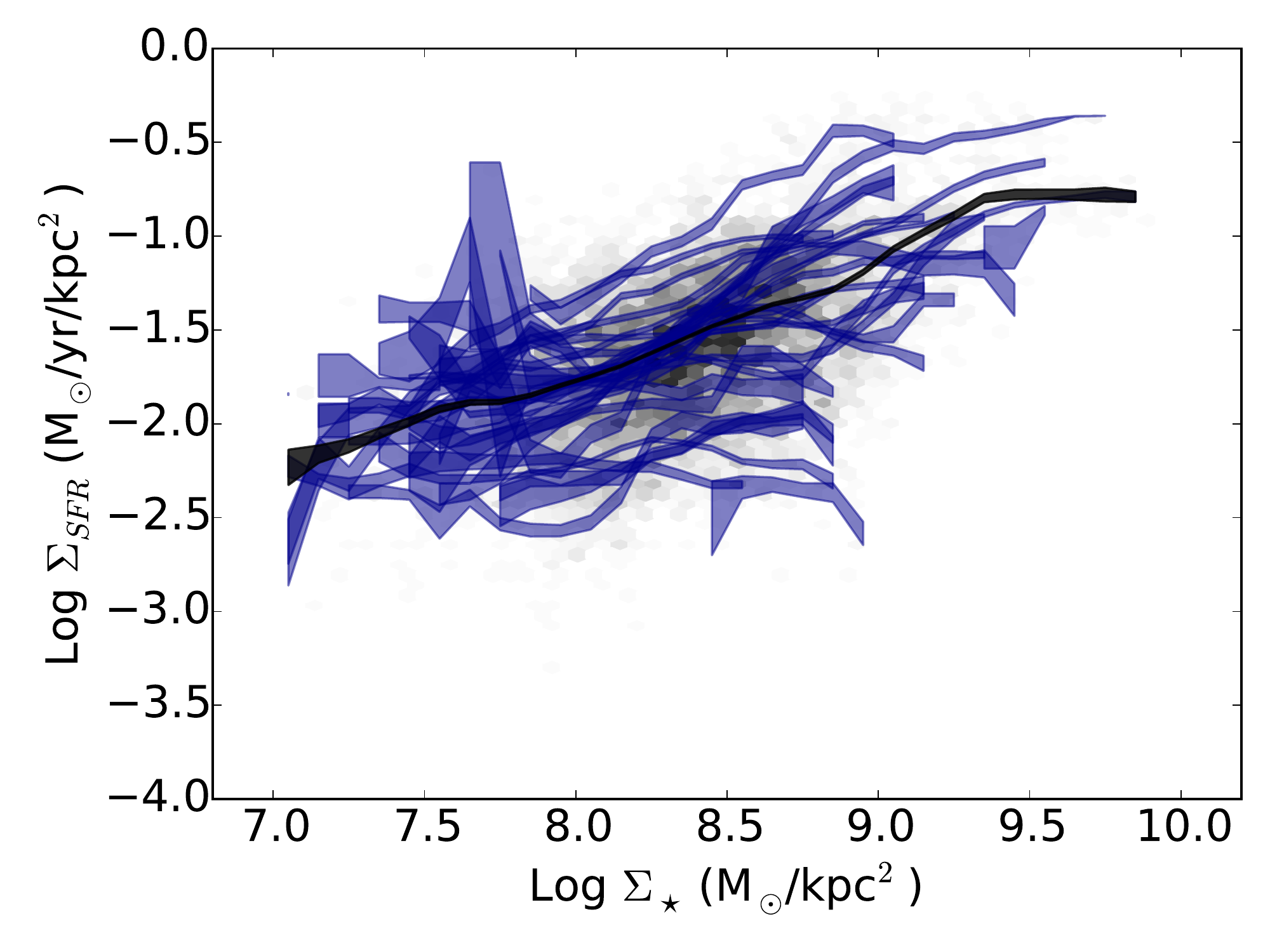}
	\includegraphics[width=5.5cm]{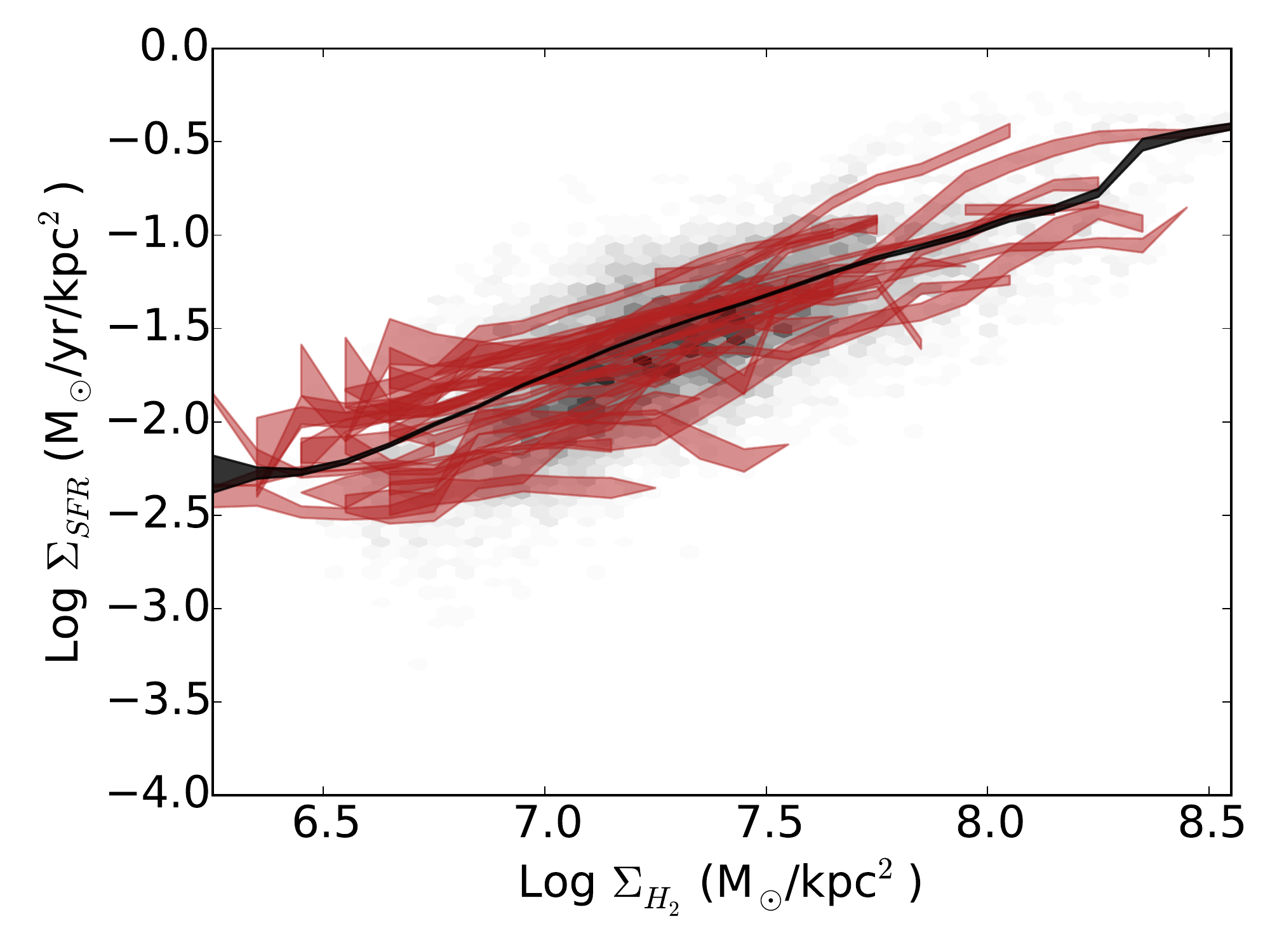}
	\includegraphics[width=5.5cm]{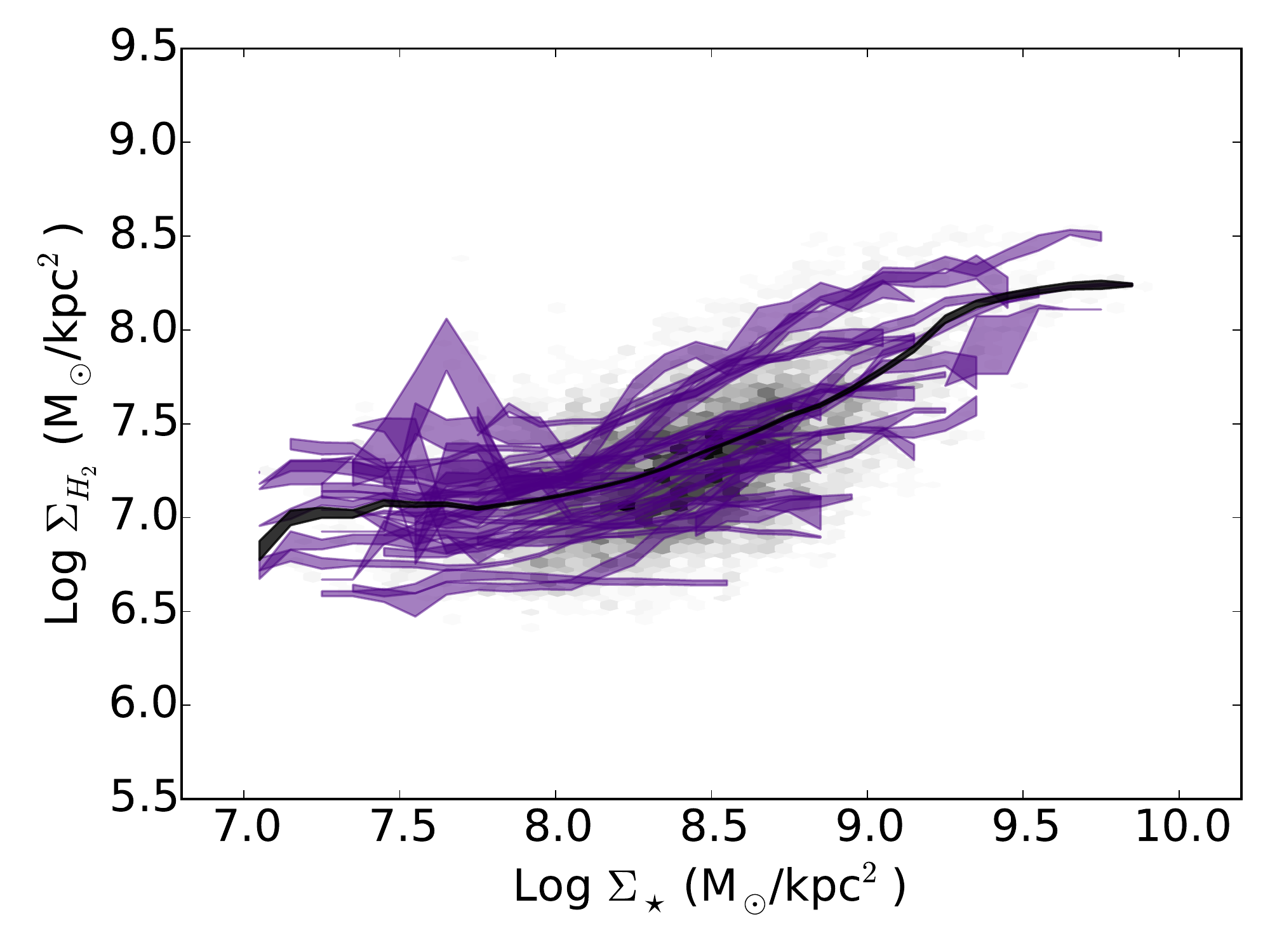}
        \caption{As for Fig. \ref{curve_offsets}, but using a metallicity dependent $\alpha_{CO}$.  Only 26/28 galaxies in our sample have at least 20 spaxels that pass our star-forming selection cut (which includes S/N thresholds in both the optical and CO data) and have robust metallicity measurements. The grey scale in the background of each panel shows the ensemble of $\sim$10,000 spaxels over all 26 galaxies.}
    \label{curve_offsets_met}
\end{figure*}

We first consider the impact of the metallicity dependent  $\alpha_{CO}$ on the ensemble relations shown in Fig. \ref{ensemble_fig}.  The rSFMS is nominally unaffected by the choice of conversion factor, since it does not involve \sigh2.  However, since we have reduced our spaxel sample to those for which we can determine a metallicity, the slope of the rSFMS used in the $\alpha_{CO}$ tests is flattened slightly to 0.61 and 1.17 for the OLS and ODR fits, respectively.

The two relations where we might expect to see a direct impact (in addition to the effect of sample selection) of a variable $\alpha_{CO}$ are the rSK relation and the rMGMS.  We find that the slope of the rSK relation is almost unaffected by the adoption of equation 1.  The OLS and ODR fits to the rSK relation with the metallicity dependent conversion factors give slope values of 0.92 and  1.27, respectively. However, a significant change is found for the slope of the ensemble rMGMS, whose value decreases to 0.86 and 0.56 for the ODR and OLS methods, respectively.  The reason for this significant flattening in the slope of the rMGMS is the presence of a resolved mass-metallicity relation (e.g. Rosales-Ortega et al. 2012; Barrera-Ballesteros et al. 2016; Ellison et al. 2018), whereby O/H increases with increasing \sigstar.  A metallicity dependent  $\alpha_{CO}$ therefore results in, broadly speaking, a \sigstar\ dependent  $\alpha_{CO}$.  Consequently, \sigh2\ tends to be slightly larger for low \sigstar\ and slightly lower for high \sigstar, resulting in a flatter gradient for the rMGMS.  There is almost no correlation between \sigsfr\ and the change in \sigh2\ using a metallicity dependent $\alpha_{CO}$, so there is no systematic change in the rSK relation slope.  When using the same sample of 10,562 spaxels with metallicity dependent $\alpha_{CO}$ conversion factors applied, we also still find that the rSFMS has the largest scatter of the three star formation relations.

In Fig. \ref{curve_offsets_met} we reproduce Fig. \ref{curve_offsets}, but with the metallicity dependendent $\alpha_{CO}$ adopted.  We still require that a galaxy has at least 20 star forming spaxels that pass all our selection criteria (including S/N in both the optical and CO data).  Given that our spaxel sample is reduced from $\sim$15,000 to $\sim$10,000 spaxels (due to the additional requirement that a robust metallicity can be computed), we find that only 26/28 galaxies in our original sample fulfill the 20 spaxel requirement.  Nonetheless, it is clear from Fig. \ref{curve_offsets_met} that the galaxy-to-galaxy diversity in all three star formation scaling relations persists when a metallicity dependent conversion factor is adopted, and that this diversity accounts for much of the scatter in the ensemble relation (grey scale in each panel).  The scatter in the curve offsets (shown for fixed $\alpha_{CO}$ in Fig. \ref{curve_offset_hist}) are found to be: $\sigma$($\Delta$rMGMS) = 0.25 dex, $\sigma$($\Delta$rSK) = 0.21 dex, $\sigma$($\Delta$rSFMS) = 0.35 dex.  The metallicity dependendent $\alpha_{CO}$ has therefore slightly increased the galaxy-to-galaxy variation in the rMGMS (originally, $\sigma$($\Delta$rMGMS) = 0.18 dex), but reduced the scatter in the rSK relation (originally, $\sigma$($\Delta$rMGMS) = 0.26 dex).  The galaxy-to-galaxy variation remains the largest in the rSFMS.

\begin{figure}
	\includegraphics[width=9cm]{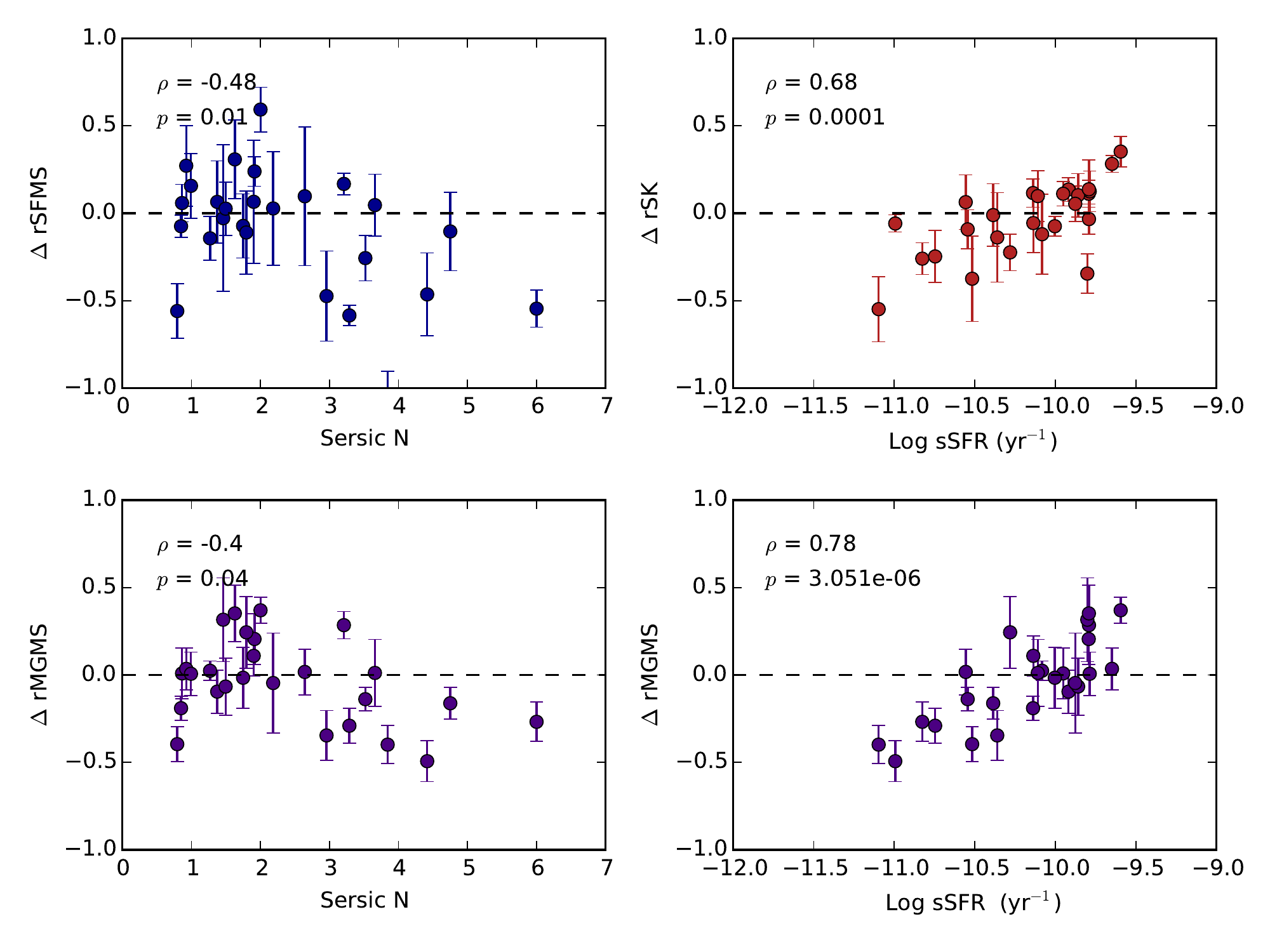}
        \caption{Significant ($>2\sigma$) correlations identified between the offset of a given star formation scaling law and various global properties when a metallicity dependent conversion factor is used.  The Pearson correlation coefficient ($\rho$) and $p$-value are given in the top left of each panel. }
    \label{stuff_met}
\end{figure}

Repeating the correlation tests shown in Figs. \ref{SFMS_stuff} to \ref{MGMS_stuff} with a metallicity dependent converstion factor, we find that the rSFMS offset still anti-correlates significantly with Sersic N ($\rho=-0.48$, $p$=0.01), although not as strongly as was found with a fixed $\alpha_{CO}$. The trend with mass also drops in strength and is no longer significant ($\rho=-0.32$, $p$=0.11).  In Fig. \ref{KS_stuff} we found significant ($>2\sigma$) correlations or anti-correlations between $\Delta$rSK and stellar mass, Sersic N and sSFR.  The adoption of a metallicity dependent $\alpha_{CO}$ weakens these correlations and the only one that remains significant with at least 2$\sigma$ significance is that between $\Delta$rSK and sSFR ($\rho=0.69$, $p=1\times10^{-4}$).  In Fig. \ref{MGMS_stuff} we found that a galaxy's offset from the average rMGMS showed a significant correlation only with its global sSFR.  Contrary to the weaker correlations found for $\Delta$rSK when a metallicity dependent conversion factor is adopted, we find that the correlation between $\Delta$rMGMS and sSFR is greatly strengthened when using a variable $\alpha_{CO}$: $\rho=0.78$, $p=3\times10^{-6}$.  The adoption of a metallicity dependent conversion factor also reveals an anti-correlation between $\Delta$rMGMS and Sersic N ($\rho=-0.4$, $p=0.04$) that was not significant for a fixed value of $\alpha_{CO}$.  For conciseness, we have not reproduced Figs \ref{SFMS_stuff} to \ref{MGMS_stuff} in full for the metallicity dependent conversion factor, but show just the correlations found at $>$2$\sigma$ significance in Fig. \ref{stuff_met}.

In summary, we find that using a metallicity dependent $\alpha_{CO}$ still leads to star formation scaling laws with significant galaxy-to-galaxy variation and that the rSFMS has both the largest scatter in the ensemble relation of all spaxels, and the largest galaxy-to-galaxy scatter.  We find significant ($>2\sigma$) trends between $\Delta$rSFMS and Sersic N (anti-correlation), between $\Delta$rSK and sSFR (correlation) and between $\Delta$rMGMS and sSFR (correlation) and Sersic N (anti-correlation).

\end{document}